\documentclass[reprint,superscriptaddress,amsmath,amssymb,aps,prx,longbibliography]{revtex4-1}

\usepackage{graphicx}
\usepackage{dcolumn}
\usepackage{bm}
\usepackage[colorlinks=true,urlcolor=blue,linkcolor=blue,citecolor=blue]{hyperref}
\usepackage[normalem]{ulem}
\usepackage{tikz}
\usetikzlibrary{quantikz}
\usepackage{amsmath}
\usepackage{amssymb}
\usepackage{bbold}
\DeclareMathOperator{\Tr}{Tr}
\usepackage{enumitem}
\usepackage[all]{hypcap} 
\usepackage{braket}

\newcommand{\customref}[2]{\hyperref[#1]{\ref*{#1}#2}}

\usepackage{xcolor}

\definecolor{Ured}{HTML}{FF5C5C}
\definecolor{Ublue}{HTML}{ADD8E6}
\definecolor{Ugreen}{HTML}{198a11}

\newcommand{\xuparrow}[1]{%
  {\left\uparrow\vbox to #1{}\right.\kern-\nulldelimiterspace}
}

\renewcommand{\vec}[1]{\boldsymbol{#1}}

\graphicspath{{figures}}

\begin{document}

%\title{Thermalization phase diagram in family of Floquet circuits}
\title{Absence of localization in weakly interacting Floquet circuits}

\author{Dominik Hahn}
\email{hahn@pks.mpg.de}
\affiliation{Max Planck Institute for the Physics of Complex Systems, N\"othnitzer Stra{\ss}e~38, 01187-Dresden, Germany}

\author{Luis Colmenarez}
\email{colmenarez@physik.rwth-aachen.de}
\affiliation{Max Planck Institute for the Physics of Complex Systems, N\"othnitzer Stra{\ss}e~38, 01187-Dresden, Germany}
\affiliation{Institute for Quantum Information, RWTH Aachen University, 52056 Aachen, Germany}
\affiliation{Institute for Theoretical Nanoelectronics (PGI-2), Forschungszentrum Jülich, 52428 Jülich, Germany}

\date{\today}%

\begin{abstract}
We present a family of Floquet circuits that can interpolate between non-interacting qubits, free propagation, generic interacting, and dual-unitary dynamics. We identify the operator entanglement entropy of the two-qubit gate as a good quantitative measure of the interaction strength. We test the persistence of localization in the vicinity of the non-interacting point by probing spectral statistics, decay of autocorrelators, and measuring entanglement growth. The finite-size analysis suggests that the many-body localized regime does not persist in the thermodynamic limit. Instead, our results are compatible with an integrability-breaking phenomenon. 
\end{abstract}
\maketitle

\section{Introduction}

Thermalization of quantum many-body systems has been an active topic of research in recent years \cite{nandkishore_many-body_2015,abanin_colloquium_2019,ueda_quantum_2020}.  It is now known that generic interacting systems reach an equilibrium state that can be described by a statistical mechanics ensemble \cite{deutsch_quantum_1991,srednicki_chaos_1994,rigol_thermalization_2008}. Such systems show quantum chaotic dynamics, which encompasses several dynamical and static traits \cite{heller_quantum_2001,dalessio_quantum_2016}. There are two non-generic cases where the dynamics is not expected to be chaotic: (i) non-interacting systems can show free propagation or get localized in the presence of disorder, the latest being called Anderson localization \cite{anderson_absence_1958}. (ii) Interacting integrable systems whose extensive number of conserved quantities constrained their dynamics, giving rise to a special type of thermalization \cite{vidmar_generalized_2016}. When integrability is perturbed in an extensive manner, quantum chaos is expected to set in the long run and the system thermalizes \cite{brenes_eigenstate_2020,rigol_fundamental_2016,essler_quench_2016,dalessio_quantum_2016,rigol_relaxation_2007}. However, in recent years, it was shown that strong disorder potentials may cause the break down of thermalization, the so-called many-body localization (MBL) \cite{basko_metalinsulator_2006,gornyi_interacting_2005,oganesyan_localization_2007}. This is described as an ergodicity-breaking transition at finite interaction strength and disorder. In this regime, quasi-local conserved quantities \cite{huse_phenomenology_2014,serbyn_local_2013} are stabilized, leading to \emph{emergent} integrability out of an initially quantum chaotic system. There are lots of evidence of MBL in systems of moderate size \cite{schreiber_observation_2015,luitz_many-body_2015}, but recently, the fate of MBL in the thermodynamic limit has been put into question \cite{suntajs_quantum_2020,sels_thermalization_2023,sels_dynamical_2021,kiefer-emmanouilidis_slow_2021} and spark debates on the scaling of the critical disorder and possible de-localization mechanisms \cite{morningstar_avalanches_2022,sels_bath-induced_2022,crowley_avalanche_2020,sierant_challenges_2022,sierant_thouless_2020,panda_can_2020,crowley_constructive_2022}.

An interesting question arises at the intersection of the dynamical regimes exposed above: how stable are Anderson localized systems to small interactions? On the one hand, any small interactions are expected to drive the system towards thermal equilibrium, on the other hand, disorder may overcome the interactions and the system becomes MBL. In Ref.~\cite{laflorencie_topological_2022} the authors studied a model where the MBL persists at small enough interactions. The critical value of the localization length is the one expected from the avalanche mechanism \cite{de_roeck_many-body_2017,thiery_many-body_2018}. In Ref.~\cite{krajewski_restoring_2022} it is shown that small interactions are not effectively perturbing the Anderson localized orbitals when the disorder strength is large. In this work, we study the localization and delocalization transition in a maximally localized system subjected to small and medium-strength interactions. 

 Most of the studies that address the strong disorder and weak interaction limit focus on Hamiltonian systems so far. Here, we present a family of Floquet circuits that can interpolate between non-interacting qubits and strongly interacting systems. In this work, we carefully investigate the finite-size scaling of the putative localization-delocalization transition by analyzing various spectral and dynamical quantities\cite{huse_phenomenology_2014,luitz_many-body_2015,bardarson_unbounded_2012,serbyn_spectral_2016,ray_drive-induced_2018,lazarides_fate_2015,oganesyan_localization_2007,znidaric_many-body_2008,sierant_stability_2023}.  Our results suggest that there exists no MBL phase for our system in the thermodynamic limit. Instead, our numerics suggest a finite-size crossover, reminiscent of integrability-breaking phenomena~\cite{bulchandani_onset_2022} in clean systems.

The structure of the paper is as follows: In Sec.~\ref{sec:model}, we introduce our Floquet model and the quantities we use to study the onset of thermalization. In Sec.~\ref{sec:results}, we present our results for various commonly studied quantities in the field, together with a finite-size scaling analysis. Finally in Sec.~\ref{sec:discussion} we discuss the implications and perspectives of our work. 

\section{Model and observables}\label{sec:model}
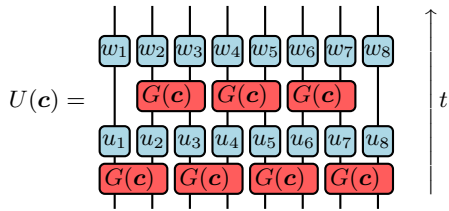
\begin{figure}[t!]
    \centering
    %\documentclass{standalone}
%\usepackage{tikz}
%\begin{document}

\begin{equation*}
U(\vec{c}) = \begin{tikzpicture}[baseline=(current bounding box.center), scale=1]
        \foreach \x in {0,0.5,1,1.5,2,2.5,3,3.5}
        {
            \draw[thick] (\x, -0.6) -- (\x, 2.1);
        }

        \foreach \x/\name in {0/1,0.5/2,1/3,1.5/4,2/5,2.5/6,3/7,3.5/8}
        {
            \draw[thick, fill=Ublue,rounded corners=2pt] (\x-0.2,0.1) rectangle +(0.4,0.4); 
            \draw (\x,0.3) node {$u_{\name}$}; 
        }

        \foreach \x/\name in {0/1,0.5/2,1/3,1.5/4,2/5,2.5/6,3/7,3.5/8}
        {
            \draw[thick, fill=Ublue,rounded corners=2pt] (\x-0.2,1.3) rectangle +(0.4,0.4); 
            \draw (\x,1.5) node {$w_{\name}$}; 
        }

        \foreach \x/\y/\name in {0/-0.4/1,1/-0.4/3,2/-0.4/5,3/-0.4/6}
        {
            \draw[thick, fill=Ured,rounded corners=2pt] (\x-0.2,\y) rectangle +(0.9,0.4); 
            \draw (\x+0.2,\y+0.2) node {$G(\vec{c})$}; 
        }

        \foreach \x/\y/\name in {0.5/0.7/1,1.5/0.7/3,2.5/0.7/5}
        {
            \draw[thick, fill=Ured,rounded corners=2pt] (\x-0.2,\y) rectangle +(0.9,0.4); 
            \draw (\x+0.2,\y+0.2) node {$G(\vec{c})$}; 
        }

    \end{tikzpicture}
    \hspace{0.3cm}
    \xuparrow{1.4cm}
    t
\end{equation*} 
    \caption{Diagrammatic notation of the unitary defined in Eq.~\eqref{eq:circuit}. The single-qubit gates $u_i$ and $w_i$ introduce spatial disorder to the system, while the interaction between neighboring qubit is mediated by the two-qubit gate $G(\mathbf{c})$.}
    \label{fig:circuit}
\end{figure}
Before we present our results, we provide technical details about our model and different quantities to detect signatures of the MBL regime.

In the following work, we study a family of  Floquet models $U(\vec{c})$ shown in Fig.~\ref{fig:circuit}, defined by the three parameters $\vec{c}=(c_1,c_2,c_3)$.

The single-qubit unitaries $u_i$ and $v_i$. are independently drawn from the Haar measure and introduce the spatial disorder of the model.
The two-qubit gates $G(\vec{c})$ are determined by three parameters $\vec{c}=(c_1,c_2,c_3)$ and are fixed for the entire circuit. The two-qubit unitary $G(\vec{c})$ is defined in Sec.~\ref{sec:circuit} and it is shown there how it naturally appears in a classification of two-qubit gates.

After that, the family of Floquet models is introduced in Sec.~\ref{sec:floquet_circuit}, together with a description of special parameter choices $\vec{c}$ into the literature. Furthermore, we present a possible characterization for the ``interaction strength" of two-qubit gates in terms of the gate operator entanglement in Sec.~\ref{sec:two-qubit}.
This section is closed with a description of the various measures used in this work to characterize the finite-size scaling of the localization-delocalization transition in Sec.~\ref{sec:level statistics} and Sec.~\ref{sec:auto_corr}.

\subsection{Classification of two-qubit gates}\label{sec:circuit}

For any two-qubit gate $U_2$, there exist general single-qubit gates $u_1,\, u_2, v_1$ and $v_2$ and a two-qubit gate of the form (with $\vec{c}=(c_1,c_2,c_3)$)
\begin{equation}
G(\vec{c}) = \exp\left[-i\dfrac{\pi}{2}\left(c_1\sigma^x \otimes \sigma^x+c_2\sigma^y \otimes \sigma^y+c_3\sigma^z \otimes \sigma^z\right)\right],
\label{eq:twositegate_local_rotation}
\end{equation}
such that it can be decomposed as ~\cite{kraus_optimal_2001}
\begin{equation}\label{eq:two_qubit_gate}
 U_2 = (u_1 \otimes u_2 ) G(\vec{c}) (v_1 \otimes v_2 ) ,
\end{equation}
Here $\sigma^{\alpha}$ with $\alpha=x,y,z$ are the Pauli operators acting on each qubit.

Restricting to the subset
${0.5\geq c_1\geq c_2 \geq c_3\geq 0}$, this allows for a classification of all two-qubit gates in terms of the vector $\vec{c}$:
Two gates which are described by the same vector $\vec{c}$ are equivalent to each other up to single-qubit gates~\cite{balakrishnan_measures_2011,balakrishnan_entangling_2010}. 

\begin{figure}[t!]
    \centering
    \centering
    \includegraphics[width=0.7\columnwidth]{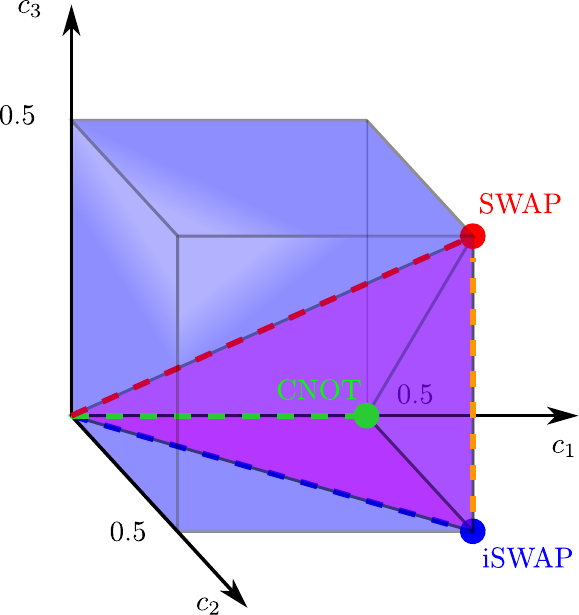}
    \caption{Parametrization of different equivalence classes for two-qubit gates: Each two-qubit gate is uniquely determined up to single-qubit unitaries by three parameters   ${0\leq c_3\leq c_2\leq c_1\leq 0.5}$~(purple tetrahedron) that form the Weyl chamber. Colored points indicate common gates. the dashed lines show the corresponding families of two-qubit gates defined in Sec.~\ref{sec:floquet_circuit}.}
    \label{fig:weyl_chamber}
\end{figure}

The space enclosed by the independent set of $\mathbf{c}$ is known as the Weyl chamber \cite{balakrishnan_characterizing_2009,zhang_geometric_2003,makhlin_nonlocal_2002} and indicated by the purple tetrahedron in Fig.~\ref{fig:weyl_chamber}. There are a few special points and lines \cite{balakrishnan_measures_2011}(see Fig.~\ref{fig:weyl_chamber}): 
\begin{enumerate}[label=(\roman*)]
    \item $c_1=c_2=c_3=0$ corresponds to independent single-qubit rotations.
    \item $c_1=c_2=c_3=0.5$ is the equivalence class of the SWAP gate up to single qubit rotations. 
    \item $c_1=c_2=0.5$ and $c_3\geq0$ denote the family  of dual-unitary gates~(shown as an orange line in Fig.~\ref{fig:weyl_chamber}).
    \item $c_1=0.5$ and $c_2=c_3=0$ is the CNOT gate up to single qubit rotations.
\end{enumerate}    
Since any two-qubit gate corresponds to a specific point of the Weyl chamber, the entanglement properties of each gate Eq.~\eqref{eq:two_qubit_gate} are uniquely determined by the vector ${\vec{c}=(c_1,c_2,c_3)}$ \cite{balakrishnan_entangling_2010,balakrishnan_operator-schmidt_2011,balakrishnan_measures_2011}, as will be illustrated in Sec.~\ref{sec:two-qubit}.

\subsection{Floquet circuit model}\label{sec:floquet_circuit}

The characterization of two-qubit unitaries by means of the gate $G(\mathbf{c})$ allows us to introduce the following family of Floquet circuits:
\begin{align}\label{eq:circuit}
    U(\vec{c})=\prod_{i=0}^{L-1} w_i \prod_{k=0}^{L/2-2} G(\vec{c})_{2k+1,2k+2}\prod_{i=0}^{L-1} u_i \prod_{k=0}^{L/2-1} G(\vec{c})_{2k,2k+1}
\end{align}
A diagrammatic representation of this circuit is shown in Fig.~\ref{fig:circuit}:
The gates $u_{i}$ and $w_i$ are single qubit rotations drawn from the Haar measure. They act as a spatial disorder. The gate $G(\vec{c})$ is defined in equation Eq.~\eqref{eq:twositegate_local_rotation}. All bonds have the same set of parameters ${\vec{c}=(c_1,c_2,c_3)}$. The Floquet dynamics is given by repeatedly applying $U$ Eq. \eqref{eq:circuit}, thus the Floquet circuit ensemble is determined by the vector $\vec{c}$. 

A few special cases of this model have been already studied in the literature:
The case of space-time dual unitary circuits $c_1=c_2=0.5$, $c_3\geq0$ (see orange dashed line in Fig.~\ref{fig:weyl_chamber}) has been extensively studied \cite{bertini_exact_2019,fisher_random_2023}. These circuits are quantum chaotic despite being exactly solvable, which makes them special in the study of thermalization \cite{piroli_exact_2020,bertini_operator_2020,bertini_operator_2020-1,claeys_ergodic_2021,aravinda_dual-unitary_2021,bertini_random_2021,bertini_exact_2018,borsi_construction_2022}. For instance, they are shown to saturate bounds on information scrambling \cite{aravinda_dual-unitary_2021,bertini_entanglement_2019, bertini_scrambling_2020,ZhouMaximal2022}. 
A random version of this circuit - different single qubit gates at each time step - was studied in Ref.~\cite{bensa_fastest_2021}, the authors found that the fastest scrambler quantum circuit, i.e., highest entanglement rate production, are random circuits with $c_1=c_2=0.5$ and $c_3\geq0$.

The vicinity of the non-interacting point $c_1=c_2=c_3=0$ is less explored. A natural question is whether the system gets many-body localized \cite{abanin_colloquium_2019,alet_many-body_2018} for small finite values of the coefficients $(c_1,c_2,c_3)$. In principle, there are multiple possible choices for $\vec{c}$ that can be studied starting from the origin. We focus on three lines:

\begin{enumerate}[label=(\roman*)]
\item SWAP line: it is determined by $c_1=c_2=c_3$ with $0<c_1<0.5$ (see Fig~\ref{fig:weyl_chamber}). It interpolates between single qubit rotations ($c_1=0$) and the SWAP gate ($c_1=0.5$). 
\item CNOT line: denoted by $c_2=c_3=0$ and $0<c_1<0.5$, this line ends on a CNOT gate-based circuit. 
\item The imaginary SWAP~(iSWAP)~\cite{Foxen2020Demonstrating} line given by $c_3=0$, $c_1=c_2$ and $0<c_1<0.5$ that ends on the imaginary-SWAP gate $c_1=c_2=0.5$. 
\end{enumerate}

In Ref.~\cite{garratt_many-body_2021} the authors report MBL for small coefficients on the SWAP line. Besides the three lines mentioned above, we study a subset $\mathbf{c}=(c_1,c_2,c_3)$ with $c_1,c_2 \in (0.02,0.28)$ and $c_3\in (0.0,0.18)$ in order to test the generality of the chosen lines and possible links between long-time dynamics and two-qubit gate invariants. 

\subsection{Two-qubit gate entanglement}\label{sec:two-qubit}
As mentioned in previous sections, a possible quantity to characterize the ``interaction strength" in the introduced class of Floquet models is by means of their operator Schmidt decomposition \cite{balakrishnan_operator-schmidt_2011}:

\begin{eqnarray}\label{eq:schmidt_decomposition}
U_2 = \sum_{l=1}^{4} \lambda_l (A_l\otimes B_l),
\end{eqnarray}

where $A_l$ and $B_l$ are orthonormal operators for the corresponding single-qubit spaces. This is analogous to the Schmidt decomposition of states. The Schmidt coefficients $\lambda_l$ are normalized, i.e. $\sum_{l=1}^{4} \lambda_l=1$. It turns out that two-qubit gates Eq.~\eqref{eq:two_qubit_gate} that are different only in single-qubit rotations will have the same set of Schmidt coefficients \cite{balakrishnan_operator-schmidt_2011}. Henceforth, the set $(c_1,c_2,c_3)$ uniquely determines any function of the Schmidt coefficients.  In the following, We use the second Renyi entropy \cite{balakrishnan_operator-schmidt_2011,balakrishnan_measures_2011} which allows us to define the operator entanglement per gate as:

\begin{eqnarray}\label{eq:schmidt_decomposition}
s(\vec{c}) = - \log\left( \sum_{l=1}^{4} \lambda^2_l \right).
\end{eqnarray}
There are a few remarks with respect to the operator entanglement for some regions in the Weyl chamber. First, the line $c_1=c_2=0.5$, $c_3<0.5$ has the largest possible  $s(\vec{c})=2\log2$ \cite{balakrishnan_operator-schmidt_2011}. They are the building block of so-called dual-unitary circuits~\cite{Bertini2019Entanglement,bertini_scrambling_2020,Akila_2016}, which exhibit the fastest scrambling in the family of random circuits \cite{bensa_fastest_2021, ZhouMaximal2022}. Recently, it was shown that after perturbing the dual unitary point the gate operator entanglement plays a crucial role in recovering the more generic quantum chaotic behavior \cite{rampp_dual_2023}. This motivates our choice to study dynamical signatures of the localization-delocalization transition as a function of the operator entanglement of the two-qubit gate they consist of.

%It is important to note that the operator entanglement entropy measures the operator distribution in operator space, not to get confused with the entangling power \cite{zanardi_entangling_2000} which is a measure of entanglement in the resulting state after the gate is applied to a product state.

The operator entanglement at any point of the Weyl chamber is given by \cite{balakrishnan_operator-schmidt_2011}:

\begin{eqnarray}\label{eq:opee_cs}
s(\vec{c}) = -\log(P(\vec{c})/32),
\end{eqnarray}
with
\begin{eqnarray}\label{eq:opee_cs_2}
& & P(\vec{c}) = 14+4\cos(2 \pi c_1)+4\cos(2 \pi c_2)+4\cos(2 \pi c_3) + \nonumber \\
           && \cos(2 \pi (c_1-c_2))+\cos(2 \pi (c_1+c_2))+\cos(2 \pi(c_1-c_3)) + \nonumber \\
           && \cos(2 \pi(c_1+c_3))+\cos(2 \pi(c_2+c_3))+\cos(2 \pi(c_2-c_3)).
\end{eqnarray}
%
%In this order we study the long-time many-body properties - i.e., scrambling, chaos and localization - of the circuit as a function of the two-qubit gate operator entanglement. Other works focused on entangling power and operator entanglement of the full circuit after a fixed number of iterations \cite{jonnadula_entanglement_2020,jonnadula_impact_2017,bensa_fastest_2021}.  

\subsection{Level statistics and eigenstate entanglement entropy}\label{sec:level statistics}

\begin{figure}
    \centering
    \includegraphics{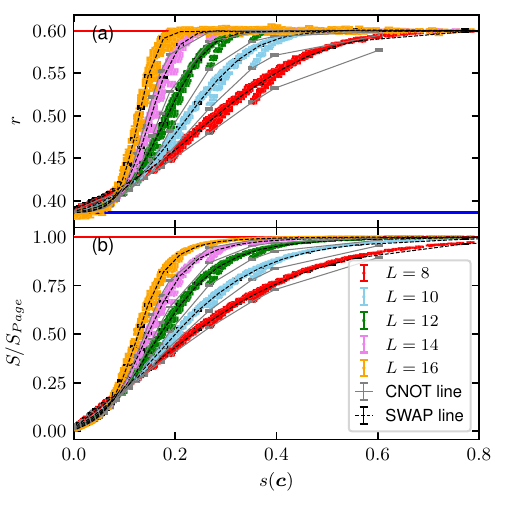}
    \caption{Gap ratio (upper panel) and eigenstate entanglement entropy (lower panel) for system sizes $L=8,10,12,14,16$ and combination of values: $c_1, c_2  = 0.02,0.04,0.06,...,0.2,0.24,0.28$ and $c_3=0,0.02,0.04,...,0.18$. SWAP line (black dashed) and CNOT line (grey continuous) are also shown for comparison. Error bars are $68\%$ confidence interval.}
    \label{fig:gap_ratio_EE_all}
\end{figure}

\begin{figure*}
    %\centering
    \includegraphics[width=\textwidth]{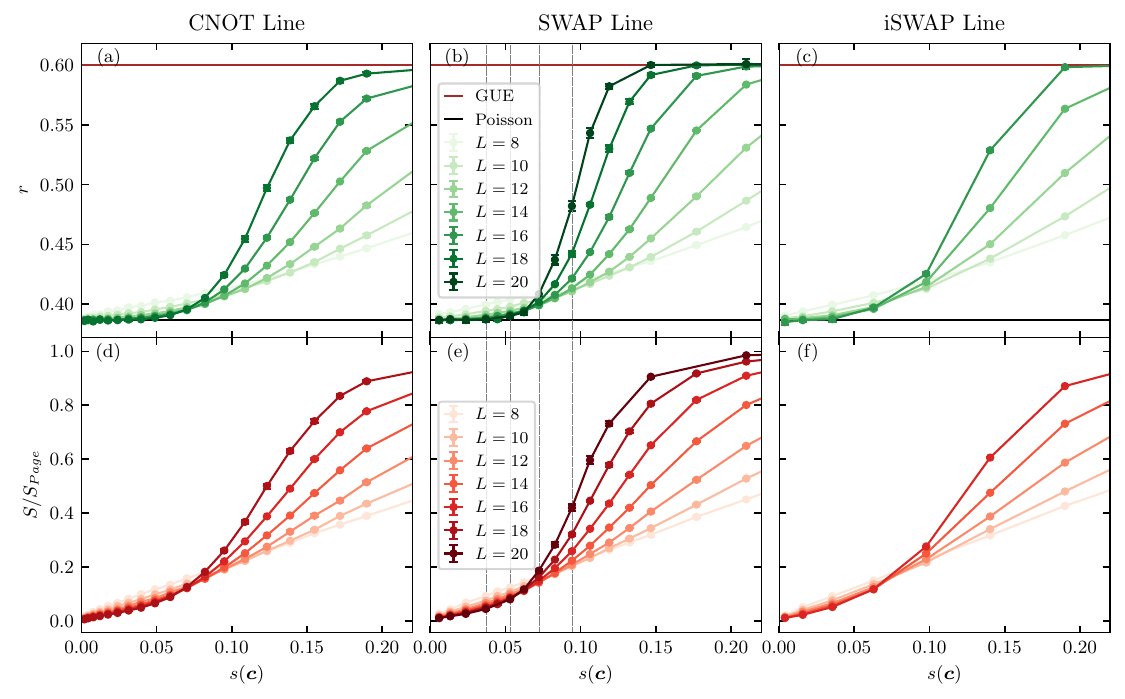}
    \caption{Upper row (\textbf{a,b,c}): Gap ratio as function of the two-qubit-gate operator entanglement entropy for system sizes $L=8,10,12,14,16$ (in addition CNOT line has $L=18$ and SWAP line has $L=18,20$). Lower row (\textbf{d,e,f}): Average entanglement entropy normalized by the Page value $S_{\text{Page}} = 0.5(L\log2-1)$ for the same system sizes. Right column: CNOT line denoted by $c_1\in [0.02,0.40]$, $c_2=c_3=0$.  Middle column: SWAP line denoted as $c_1=c_2=c_3$ with $c_1\in [0.02,0.40]$. Left column: iSWAP line denoted by $c_2=c_1$, $c_3=0$ and $c_1\in [0.02,0.28]$. Error bars (too small for this scale) are $68\%$ confidence interval. Grey vertical lines are the $s(\vec{c})$ for which dynamics is shown in Sec.~\ref{sec:dynamics}. In all three cases, the operator entanglement entropy $s(\vec{c})$ indicates an MBL-to-thermal crossover for both gap ratio and eigenstate entanglement entropy.}
   \label{fig:gapratio_and_EE}
\end{figure*}

The long-time dynamics of Floquet circuits can be probed by the spectral properties of the Floquet operator \cite{dalessio_quantum_2016}. In particular, we are interested in the eigenphases and eigenstates $U |n\rangle = e^{i\theta_n} |n\rangle$. The gaps between consecutive eigenphases are defined as $\delta_i = \theta_{i+1}-\theta_i$, the ratio between two consecutive gaps is denoted as:
\begin{eqnarray}\label{eq:schmidt_decomposition}
 r_i = \min\left(\delta_{i+1},\delta_i\right)/\max\left(\delta_{i+1},\delta_i\right).
\end{eqnarray}
The average gap ratio $r$ is known to serve as an order parameter for ergodicity breaking transitions \cite{pal_many-body_2010,luitz_many-body_2015}: When the Floquet dynamics leads to thermalization, the mean gap ratio is described by GUE random matrix ensemble $\overline{r}\approx0.60$ \cite{atas_distribution_2013}. In contrast, when the Floquet dynamics gets localized, i.e. MBL, the gap ratio statistics is Poissonian such that $\overline{r}=2\log2-1 \approx 0.386$ \cite{oganesyan_localization_2007,dalessio_quantum_2016}. If our Floquet model undergoes a MBL transition it should show up in the behavior of $r$ as a function of the Schmidt coefficients. 

A second diagnostic for the transition is the  structure of eigenstates of the time evolution operator:  We introduce the reduced density matrix over half of the system as  $\rho_A = \Tr_{L/2} \left(|n\rangle\langle n|\right)$. 
We probe the transition at the level of eigenstates using the half-chain entanglement entropy 
\begin{equation}\label{Eq:entropy}
S = -\Tr \rho_A \log \rho_A .
\end{equation}
In the thermal phase, the eigenstates are expected to be essentially random vectors in Hilbert space; thus their entanglement entropy is proportional to the chain length $S_{\text{Page}} = (L\log2-1)/2$, the so-called Page value \cite{page_average_1993}. On the other hand, in the localized phase the eigenstates only exhibit short-range entanglement, resulting in an area law for the entanglement entropy $S\approx const$~\cite{nandkishore_many-body_2015}. Hence the average entanglement entropy $\overline{S}$ signals the MBL transition \cite{khemani_critical_2017,luitz_many-body_2015,yu_bimodal_2016}. 

These quantities are obtained using exact diagonalization. For small system sizes $L=8,10,12$ the whole eigenspectrum is computed, while for larger system sizes $L\geq 14$ polynomial filtered diagonalization \cite{luitz_polynomial_2021} is used for extracting 100 eigenpairs of the Floquet unitary. All quantities are averaged over 3000-6000 disorder realizations (except for $L=20$ where only 500-1000 realizations are used) and all available eigenstates. 

\subsection{Quench dynamics}\label{sec:auto_corr}
Another direct way to detect localization in our system is using transport properties. A common tool is the autocorrelator~\cite{long_phenomenology_2022,schreiber_observation_2015,luitz_extended_2016,lezama_apparent_2019,lezama_equilibration_2021}:
\begin{eqnarray}
C(t) = \dfrac{1}{2^L} \Tr \left[\hat{O}(t)\hat{O}\right]=\dfrac{1}{2^L} \Tr \left[(\hat{U}^{\dagger}(t) \hat{O} \hat{U}(t)\hat{O}\right].
\end{eqnarray}

Here $O$ is a normalized observable with vanishing mean~($\Tr{\hat{O}}=0$, $\Tr{\hat{O}^2}=2^L$), and the time evolution is generated by the circuit introduced in Eq.~\eqref{eq:circuit}, i.e. $\hat{U}(t)=U(\vec{c})^t$. 
 
Since our goal is to probe scrambling caused by the entangling gates, we choose an operator $\hat{O}$ such that $C(t)=1$ for $c_1=c_2=c_3=0$.
To achieve this, consider the product of the single-site operators at $i=\lfloor L/2 \rfloor$:
\begin{align}
    R=u_i w_i
\end{align}
Since $R$ is unitary, there exists a diagonal matrix $D$ and an unitary $V$ such that
\begin{align}
    R=V^\dagger D V.
\end{align}
By choosing 
\begin{align}
\hat{O} = V^{\dagger} \sigma^z_i V
\end{align}
we obtain an autocorrelator $C(t)$ with the desired properties. It is important to note that the choice of $V$ depends on the specific disorder configuration. When the system thermalizes, the disorder-averaged autocorrelator vanishes in the long-time limit~\cite{nandkishore_many-body_2015}. In contrast, $C(t)$ is expected to converge to a non-zero value in the MBL regime~\cite{nandkishore_many-body_2015}. In summary, we obtain
\begin{align}
    \lim_{t\rightarrow \infty} C(t)=\begin{cases}
0, &\text{thermalization}\\
c>0, &\text{localization}.\\
\end{cases}
\end{align}

Another way to probe transport properties in the study of MBL is the entanglement entropy production starting from a product state $|\psi(0)\rangle=|\psi_0\rangle$ \cite{znidaric_many-body_2008,kjall_many-body_2014}. 
Analog to Sec.~\ref{sec:level statistics}, we compute the half-chain entanglement entropy $S(t)$~[cf.~Eq.\eqref{Eq:entropy}], but now for the time-evolved state $|\psi(t)\rangle$ instead. The entanglement growth rate depends on the overall dynamics: quantum chaotic systems show linear growth in time, while MBL systems exhibit logarithmic growth of entanglement \cite{znidaric_many-body_2008,bardarson_unbounded_2012}. The latest signals the existence of quasi-local integrals of motion \cite{serbyn_local_2013,huse_phenomenology_2014}. 

%Persisting logarithmic growth of entanglement and steady non-vanishing auto correlations are dynamical signatures of MBL regimes, at least for the accessible time scales. 

\section{Results}\label{sec:results}
\subsection{Gap ratio and eigenstate entanglement entropy}

\begin{figure}
    \centering
    \includegraphics{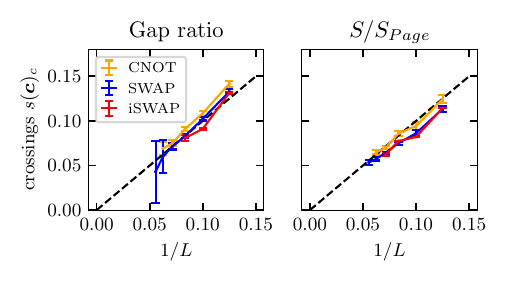}
    \caption{Crossings of the curves in Fig.~\ref{fig:gapratio_and_EE} between system size $L$ and $L+2$ for the gap ratio~(a) and eigenstate entanglement entropy~(b) along the CNOT line (orange), SWAP line (blue), iSWAP line (red). The black dashed lines show the scaling $1/L$. For accessible system sizes the trend is compatible within error bars with $s(\vec{c})_c \propto 1/L$, suggesting restoring of ergodicity in the thermodynamic limit for any finite interaction strength.}
    \label{fig:crossings}
\end{figure}

As a first check, we probe the operator entanglement per gate $s(\vec{c})$ as a unifying parameter for the interaction strength. To do so, we show the gap ratio and the eigenstate entanglement entropy as a function
of the gate entanglement entropy $s(\vec{c})$ in Fig.~\ref{fig:gap_ratio_EE_all} (a) and (b), respectively. We focus on parameters $\vec{c}(c_1,c_2,c_3)$ within the range $c_1,c_2 = [0.02,0.28]$ and $c_3 = [0.0,0.18]$. For the system size and $s(\vec{c})$ fixed, the gap ratio and half-chain entanglement almost collapse on top of the corresponding SWAP line value. This supports our motivation to choose $s(\vec{c})$ as an indicator for the interaction strength.

However, the results for the CNOT and the SWAP line lie not directly on top of each other. This difference may originate from the specific choice of the vector $\vec{c}$. On the CNOT line, two Schmidt coefficients $\lambda_l$~[cf.  Eq.~\eqref{eq:schmidt_decomposition}] are zero \cite{balakrishnan_operator-schmidt_2011}, which is not the case for any other choice of $\vec{c}$ in the Weyl chamber. Therefore the finite size behavior visible in $r$ and $S/S_{\text{Page}}$ may be affected by this choice of $\vec{c}$ (see Appendix~\ref{sec:opee_swap_cnot} for a more quantitative comparison).
\begin{figure*}
    \centering
    \includegraphics[width=\textwidth]{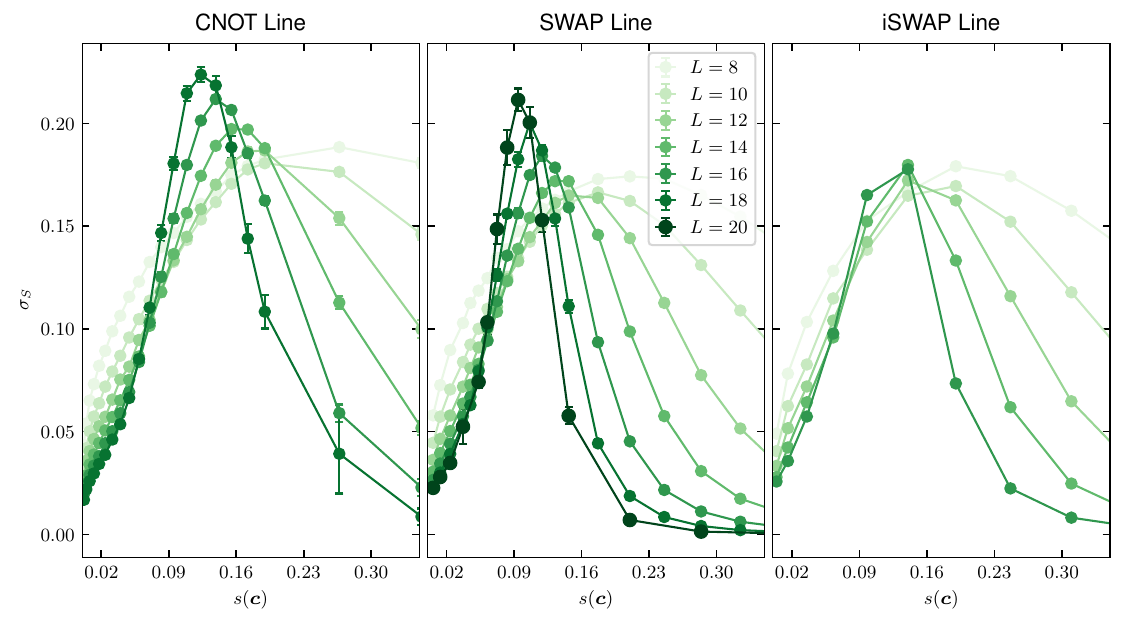}
    \caption{Entanglement entropy fluctuations $\sigma_S^2 = \overline{s^2}-\overline{s}^2$ with $s=S/S_{\text{Page}}$ as function of the two-qubit gate operator entanglement. Right column: CNOT line denoted by $c_1\in [0.02,0.40]$. Middle column: SWAP line with $c_1\in [0.02,0.40]$. Left column: iSWAP line with $c_1\in [0.02,0.28]$. Error bars (too small for this scale) are $68\%$ confidence interval. The data set is the same shown in Fig.\ref{fig:gapratio_and_EE} for $S/S_{\text{Page}}$.}
    \label{fig:std_EE}
\end{figure*}

For our further studies, we choose three parameter lines that differ in the number of non-vanishing coefficients $c_i$: the SWAP line, the iSWAP line, and the CNOT line~(cf Sec.~\ref{sec:floquet_circuit}). The gap ratio and the half-chain entanglement entropy on these lines of the Weyl chamber and systems sizes are shown in Fig.~\ref{fig:gapratio_and_EE}. In all three cases, we see a crossover between an MBL regime~(indicated by a small eigenstate entanglement entropy and an $r$-value close to the Poissonian case) for small interactions towards a thermal regime at large interactions. 

The curves of different system sizes intersect. Analogously to Ref.~\cite{morningstar_avalanches_2022,oganesyan_localization_2007,luitz_many-body_2015}, we then compare the different crossings between consecutive system sizes to check the stability of the phase in the thermodynamic limit~\cite{pal_many-body_2010}. In Fig.~\ref{fig:crossings}, the crossings of gap ratio and entanglement entropy for consecutive sizes $L$ and $L+2$ are shown. The trend of the crossing suggests a scaling $\propto 1/L$, at least for the accessible system sizes. Remarkably, finite size effects for the half-chain entanglement entropy are less pronounced in comparison to the r-value.
The trend of the data suggests that the crossings are shifting towards zero in the limit $L\rightarrow\infty$, thus ergodicity is restored at any finite interaction. However, given the smallness of the accessible system sizes, we can not rule out a change in the trend at larger sizes.

Finally, we present an analysis of the entanglement entropy fluctuations of the Floquet eigenstates.
It is known that fluctuations around the mean eigenstate entanglement entropy $\overline{S}$, probed by $\sigma_S^2 = \overline{S^2}-\overline{S}^2$, peak at the transition point \cite{luitz_many-body_2015} and thus are a good indicator to identify the delocalization-localization transition~\cite{Ponte2015Many-body,Kjaell2014Many-body}. Such fluctuations are shown in Fig.~\ref{fig:std_EE}. We see that the peak of the fluctuations is moving towards smaller $s(\vec{c})$, as has been reported in other models where MBL might be stable~\cite{yu_bimodal_2016}. Importantly, the entanglement entropy fluctuations also seem to be sensitive only to the operator entanglement $s(\vec{c})$ of the gate rather than the specific choice $(c_1,c_2,c_3)$ in the Weyl chamber.

\subsection{Single spin autocorrelation and entanglement entropy after a quench}\label{sec:dynamics}

\begin{figure}
    \centering
    \includegraphics{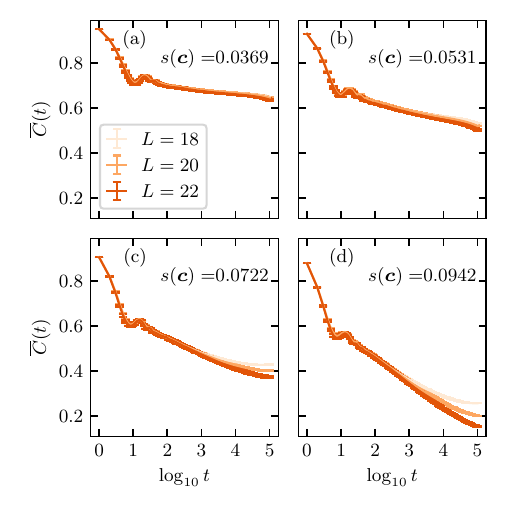}
    \caption{Single spin autocorrelation $\overline{C}(t)$ as a function of time $t$, gates $G(\vec{c})$ are chosen on the SWAP line for $c=(0.05,0.06,0.07,0.08)$ (the corresponding interaction strength $s(\vec{c})$ is shown in each panel). Disorder average is taken over 1000-4000 disorder realizations. Errors bars are $68\%$ confidence interval.}
    \label{fig:auto_correlation}
\end{figure}

\begin{figure}
    \centering
    \includegraphics{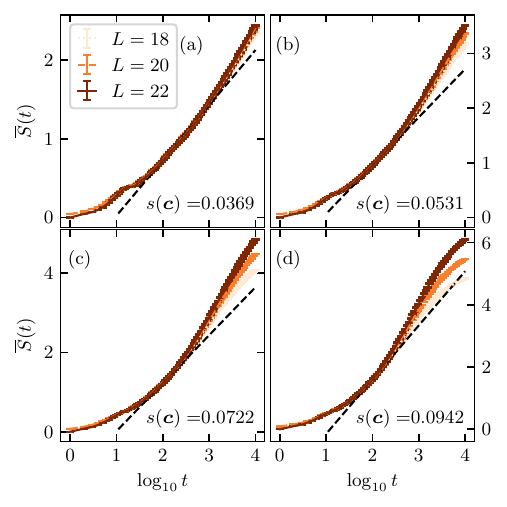}
    \caption{Half chain entanglement entropy after a quench on the SWAP line for $c=(0.05,0.06,0.07,0.08)$  (same values but translated to $s(\vec{c})$ are shown in each panel) and system sizes $L=18,20,22$. Disorder average is taken over 1000-4000 disorder realizations, error bars denote $68\%$ confidence interval.  The black dashed lines denote a fit $a \log t +b$ for comparison, with $a$ determined by the derivative of the $L=22$ curve at $t=10^3$. For large interactions $S(t)$ the logarithmic growth is proceeded by a faster entanglement entropy growth.}
    \label{fig:entang_growth}
\end{figure}

\begin{figure}
    \centering
    \includegraphics{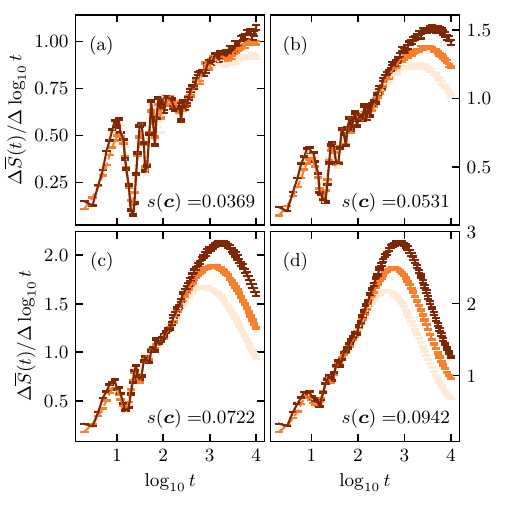}
    \caption{Numerical derivative of the half chain entanglement entropy with respect to $\log t$ after a quench on the SWAP line for $c=(0.05,0.06,0.07,0.08)$  (same data is shown in Fig.~\ref{fig:entang_growth}) and system sizes $L=18,20,22$. The entropy growth is faster than pure logarithmic for all interaction strengths.}
    \label{fig:entang_growth_der}
\end{figure}

From the previous section, we can conclude that the critical operator entanglement per gate $s(\vec{c})$  is scaling roughly as $1/L$. The largest system size for which we could extract eigenvalues and eigenvectors is $L=20$. From Fig.~\ref{fig:crossings}, we estimate the crossover region for current system sizes to be around $s(\vec{c})\sim1/20=0.05$. In this section, we explore signatures for this crossover in quench dynamics for up to $10^5$ cycles in the regime $s(\vec{c})\lesssim 0.1$ and system sizes $L\geq 18$. 

In Fig.~\ref{fig:auto_correlation}, we show the autocorrelation function introduced in Sec.~\ref{sec:auto_corr} on the SWAP line of the model. The circuit dynamics is simulated using Cirq \cite{cirq_developers_2023_8161252} that allows to reach $10^5$ cycles and up to $L=22$ qubits. 

For $s(\vec{c})\approx 0.09,0.07$ close to the crossover region, $\overline{C}(t)$ decays with a scale either logarithmically or stretched exponentially, in line with previous work on autocorrelation decay in prethermal systems~\cite{long_phenomenology_2022}. The long-time limit $\lim_{t\rightarrow \infty}C(t)$ decreases with system size, suggesting a trend towards thermalization. For smaller interactions $s(\vec{c})\approx 0.05,0.03$, our accessible time-scales are too short to draw conclusions about a drift in the long-time dynamics: we do not reach a steady state in our numerics. 

Finally, we study the entanglement entropy growth for initial product states along the SWAP line. As discussed in Sec.~\ref{sec:auto_corr}, a signature of localization is logarithmic entanglement growth. As is visible in Fig.~\ref{fig:entang_growth}, the entanglement entropy $S(t)$ is growing faster than logarithmically~(black dashed lines) at time scales $t\gtrsim 10^3$ and interactions $s(\vec{c})\gtrsim 0.07$. In order to confirm this observation, we compute the derivative of $S(t)$ with respect to $\log t$ (see Fig.~\ref{fig:entang_growth_der}). A logarithmic curve would be visible as a constant value. Instead, we see that the derivative keeps growing with system size even in the regime when the level statistics is Poissonian $s(c)=0.0369$. We conclude that the entropy growth is faster than pure logarithmic growth even for the smallest interactions. The latest is in odds with the steady logarithmic growth in MBL regimes \cite{bardarson_unbounded_2012}.

\section{Discussion}\label{sec:discussion}
In this work, we have introduced a generic Floquet circuit model that allows us to parametrically tune the interaction and keep the disorder maximal. We have identified the gate entanglement entropy $s(\vec{c})$ as a quantitative measure for the interaction. In the limit $s(\vec{c})=0$, our model reduces to a non-interacting system. Our results for various quantities suggest that the observed MBL regime for small interactions does not persist in the thermodynamic limit.

Our reachable system sizes and our investigated model are not sufficient to draw conclusions about the general fate of the MBL transition in the thermodynamic limit. Nevertheless, they suggest analyzing whether the results for other commonly studied models in the field of many-body localization are in alignment with an integrability-breaking transition~\cite{krajewski_restoring_2022,krajewski_strongly_2023} instead. 

Moreover, it is an interesting question to establish connections between this model, where both direction and strength are subjected to disorder, and models where the direction of the single-qubit unitaries is fixed. The $XXZ$ spin chain and its variants are part of the latter. Furthermore, the effect of spatial variations on the gate operator entanglement can give rise to ``slow" and ``fast" dynamical regions very much in the spirit of quantum avalanches proposed as delocalization mechanism \cite{thiery_many-body_2018,de_roeck_many-body_2017}. The investigation of the role of both types of disorder and the effects of spatial fluctuations in gate operator entanglement are both interesting venues for future research.

Apart from that, our model contains dual unitary circuits as another special case for a specific choice of parameters $\vec{c}$. This model is thus a good starting point to study the effects of breaking dual-unitarity in more detail~\cite{rampp_dual_2023}.

%While we have restricted in our case to system sizes reachable with ED, there exist methods using the density matrix renormalization group to obtain highly-excited eigenstates in the many-body regime~\cite{Khemani2016Obtaining,devakul2017obtaining}.
%This allows for obtaining accurate estimates of the thermal-MBL crossover beyond reachable system sizes with exact diagonalization. \lc{tensor network methods are highly biased towards localization, that's why people don't use them anymore for studying the transition/crossover}.

\section{Acknowledgments}

We thank Anushya Chandran, Pieter Claeys, David Long, David Luitz and Michael Rampp for inspiring discussions. L.C. gratefully acknowledges funding by the U.S. ARO Grant No. W911NF-21-1-0007. All statements of fact, opinion or conclusions contained herein are those of the authors and should not be construed as representing the official views or policies of the US Government.

\bibliography{floquet_circuit_generic,floquet_mbl_circuit,cirq_ref}

\begin{thebibliography}{85}%
\makeatletter
\providecommand \@ifxundefined [1]{%
 \@ifx{#1\undefined}
}%
\providecommand \@ifnum [1]{%
 \ifnum #1\expandafter \@firstoftwo
 \else \expandafter \@secondoftwo
 \fi
}%
\providecommand \@ifx [1]{%
 \ifx #1\expandafter \@firstoftwo
 \else \expandafter \@secondoftwo
 \fi
}%
\providecommand \natexlab [1]{#1}%
\providecommand \enquote  [1]{``#1''}%
\providecommand \bibnamefont  [1]{#1}%
\providecommand \bibfnamefont [1]{#1}%
\providecommand \citenamefont [1]{#1}%
\providecommand \href@noop [0]{\@secondoftwo}%
\providecommand \href [0]{\begingroup \@sanitize@url \@href}%
\providecommand \@href[1]{\@@startlink{#1}\@@href}%
\providecommand \@@href[1]{\endgroup#1\@@endlink}%
\providecommand \@sanitize@url [0]{\catcode `\\12\catcode `\$12\catcode
  `\&12\catcode `\#12\catcode `\^12\catcode `\_12\catcode `\%12\relax}%
\providecommand \@@startlink[1]{}%
\providecommand \@@endlink[0]{}%
\providecommand \url  [0]{\begingroup\@sanitize@url \@url }%
\providecommand \@url [1]{\endgroup\@href {#1}{\urlprefix }}%
\providecommand \urlprefix  [0]{URL }%
\providecommand \Eprint [0]{\href }%
\providecommand \doibase [0]{http://dx.doi.org/}%
\providecommand \selectlanguage [0]{\@gobble}%
\providecommand \bibinfo  [0]{\@secondoftwo}%
\providecommand \bibfield  [0]{\@secondoftwo}%
\providecommand \translation [1]{[#1]}%
\providecommand \BibitemOpen [0]{}%
\providecommand \bibitemStop [0]{}%
\providecommand \bibitemNoStop [0]{.\EOS\space}%
\providecommand \EOS [0]{\spacefactor3000\relax}%
\providecommand \BibitemShut  [1]{\csname bibitem#1\endcsname}%
\let\auto@bib@innerbib\@empty
%</preamble>
\bibitem [{\citenamefont {Nandkishore}\ and\ \citenamefont
  {Huse}(2015)}]{nandkishore_many-body_2015}%
  \BibitemOpen
  \bibfield  {author} {\bibinfo {author} {\bibfnamefont {Rahul}\ \bibnamefont
  {Nandkishore}}\ and\ \bibinfo {author} {\bibfnamefont {David~A.}\
  \bibnamefont {Huse}},\ }\bibfield  {title} {\enquote {\bibinfo {title}
  {Many-{Body} {Localization} and {Thermalization} in {Quantum} {Statistical}
  {Mechanics}},}\ }\href
  {https://doi.org/10.1146/annurev-conmatphys-031214-014726} {\bibfield
  {journal} {\bibinfo  {journal} {Annual Review of Condensed Matter Physics}\
  }\textbf {\bibinfo {volume} {6}},\ \bibinfo {pages} {15--38} (\bibinfo {year}
  {2015})}\BibitemShut {NoStop}%
\bibitem [{\citenamefont {Abanin}\ \emph {et~al.}(2019)\citenamefont {Abanin},
  \citenamefont {Altman}, \citenamefont {Bloch},\ and\ \citenamefont
  {Serbyn}}]{abanin_colloquium_2019}%
  \BibitemOpen
  \bibfield  {author} {\bibinfo {author} {\bibfnamefont {Dmitry~A.}\
  \bibnamefont {Abanin}}, \bibinfo {author} {\bibfnamefont {Ehud}\ \bibnamefont
  {Altman}}, \bibinfo {author} {\bibfnamefont {Immanuel}\ \bibnamefont
  {Bloch}}, \ and\ \bibinfo {author} {\bibfnamefont {Maksym}\ \bibnamefont
  {Serbyn}},\ }\bibfield  {title} {\enquote {\bibinfo {title} {Colloquium:
  {Many}-body localization, thermalization, and entanglement},}\ }\href
  {https://link.aps.org/doi/10.1103/RevModPhys.91.021001} {\bibfield  {journal}
  {\bibinfo  {journal} {Rev. Mod. Phys.}\ }\textbf {\bibinfo {volume} {91}},\
  \bibinfo {pages} {021001} (\bibinfo {year} {2019})}\BibitemShut {NoStop}%
\bibitem [{\citenamefont {Ueda}(2020)}]{ueda_quantum_2020}%
  \BibitemOpen
  \bibfield  {author} {\bibinfo {author} {\bibfnamefont {Masahito}\
  \bibnamefont {Ueda}},\ }\bibfield  {title} {\enquote {\bibinfo {title}
  {Quantum equilibration, thermalization and prethermalization in ultracold
  atoms},}\ }\href {https://www.nature.com/articles/s42254-020-0237-x}
  {\bibfield  {journal} {\bibinfo  {journal} {Nat Rev Phys}\ }\textbf {\bibinfo
  {volume} {2}} (\bibinfo {year} {2020})}\BibitemShut {NoStop}%
\bibitem [{\citenamefont {Deutsch}(1991)}]{deutsch_quantum_1991}%
  \BibitemOpen
  \bibfield  {author} {\bibinfo {author} {\bibfnamefont {J.~M.}\ \bibnamefont
  {Deutsch}},\ }\bibfield  {title} {\enquote {\bibinfo {title} {Quantum
  statistical mechanics in a closed system},}\ }\href
  {https://link.aps.org/doi/10.1103/PhysRevA.43.2046} {\bibfield  {journal}
  {\bibinfo  {journal} {Phys. Rev. A}\ }\textbf {\bibinfo {volume} {43}},\
  \bibinfo {pages} {2046--2049} (\bibinfo {year} {1991})}\BibitemShut {NoStop}%
\bibitem [{\citenamefont {Srednicki}(1994)}]{srednicki_chaos_1994}%
  \BibitemOpen
  \bibfield  {author} {\bibinfo {author} {\bibfnamefont {Mark}\ \bibnamefont
  {Srednicki}},\ }\bibfield  {title} {\enquote {\bibinfo {title} {Chaos and
  quantum thermalization},}\ }\href
  {https://link.aps.org/doi/10.1103/PhysRevE.50.888} {\bibfield  {journal}
  {\bibinfo  {journal} {Phys. Rev. E}\ }\textbf {\bibinfo {volume} {50}},\
  \bibinfo {pages} {888} (\bibinfo {year} {1994})}\BibitemShut {NoStop}%
\bibitem [{\citenamefont {Rigol}\ \emph {et~al.}(2008)\citenamefont {Rigol},
  \citenamefont {Dunjko},\ and\ \citenamefont
  {Olshanii}}]{rigol_thermalization_2008}%
  \BibitemOpen
  \bibfield  {author} {\bibinfo {author} {\bibfnamefont {Marcos}\ \bibnamefont
  {Rigol}}, \bibinfo {author} {\bibfnamefont {Vanja}\ \bibnamefont {Dunjko}}, \
  and\ \bibinfo {author} {\bibfnamefont {Maxim}\ \bibnamefont {Olshanii}},\
  }\bibfield  {title} {\enquote {\bibinfo {title} {Thermalization and its
  mechanism for generic isolated quantum systems},}\ }\href
  {https://www.nature.com/articles/nature06838} {\bibfield  {journal} {\bibinfo
   {journal} {Nature}\ }\textbf {\bibinfo {volume} {452}} (\bibinfo {year}
  {2008})}\BibitemShut {NoStop}%
\bibitem [{\citenamefont {Heller}(2001)}]{heller_quantum_2001}%
  \BibitemOpen
  \bibfield  {author} {\bibinfo {author} {\bibfnamefont {Eric~J.}\ \bibnamefont
  {Heller}},\ }\bibfield  {title} {\enquote {\bibinfo {title} {Quantum {Chaos}:
  {An} {Introduction}},}\ }\href {https://doi.org/10.1063/1.1349613} {\bibfield
   {journal} {\bibinfo  {journal} {Physics Today}\ }\textbf {\bibinfo {volume}
  {54}},\ \bibinfo {pages} {49--50} (\bibinfo {year} {2001})}\BibitemShut
  {NoStop}%
\bibitem [{\citenamefont {D'Alessio}\ \emph {et~al.}(2016)\citenamefont
  {D'Alessio}, \citenamefont {Kafri}, \citenamefont {Polkovnikov},\ and\
  \citenamefont {Rigol}}]{dalessio_quantum_2016}%
  \BibitemOpen
  \bibfield  {author} {\bibinfo {author} {\bibfnamefont {Luca}\ \bibnamefont
  {D'Alessio}}, \bibinfo {author} {\bibfnamefont {Yariv}\ \bibnamefont
  {Kafri}}, \bibinfo {author} {\bibfnamefont {Anatoli}\ \bibnamefont
  {Polkovnikov}}, \ and\ \bibinfo {author} {\bibfnamefont {Marcos}\
  \bibnamefont {Rigol}},\ }\bibfield  {title} {\enquote {\bibinfo {title} {From
  quantum chaos and eigenstate thermalization to statistical mechanics and
  thermodynamics},}\ }\href {https://doi.org/10.1080/00018732.2016.1198134}
  {\bibfield  {journal} {\bibinfo  {journal} {Advances in Physics}\ }\textbf
  {\bibinfo {volume} {65}} (\bibinfo {year} {2016})}\BibitemShut {NoStop}%
\bibitem [{\citenamefont {Anderson}(1958)}]{anderson_absence_1958}%
  \BibitemOpen
  \bibfield  {author} {\bibinfo {author} {\bibfnamefont {P.~W.}\ \bibnamefont
  {Anderson}},\ }\bibfield  {title} {\enquote {\bibinfo {title} {Absence of
  {Diffusion} in {Certain} {Random} {Lattices}},}\ }\href
  {https://link.aps.org/doi/10.1103/PhysRev.109.1492} {\bibfield  {journal}
  {\bibinfo  {journal} {Phys. Rev.}\ }\textbf {\bibinfo {volume} {109}}
  (\bibinfo {year} {1958})}\BibitemShut {NoStop}%
\bibitem [{\citenamefont {Vidmar}\ and\ \citenamefont
  {Rigol}(2016)}]{vidmar_generalized_2016}%
  \BibitemOpen
  \bibfield  {author} {\bibinfo {author} {\bibfnamefont {Lev}\ \bibnamefont
  {Vidmar}}\ and\ \bibinfo {author} {\bibfnamefont {Marcos}\ \bibnamefont
  {Rigol}},\ }\bibfield  {title} {\enquote {\bibinfo {title} {Generalized
  {Gibbs} ensemble in integrable lattice models},}\ }\href
  {https://dx.doi.org/10.1088/1742-5468/2016/06/064007} {\bibfield  {journal}
  {\bibinfo  {journal} {J. Stat. Mech.}\ }\textbf {\bibinfo {volume} {2016}},\
  \bibinfo {pages} {064007} (\bibinfo {year} {2016})}\BibitemShut {NoStop}%
\bibitem [{\citenamefont {Brenes}\ \emph {et~al.}(2020)\citenamefont {Brenes},
  \citenamefont {LeBlond}, \citenamefont {Goold},\ and\ \citenamefont
  {Rigol}}]{brenes_eigenstate_2020}%
  \BibitemOpen
  \bibfield  {author} {\bibinfo {author} {\bibfnamefont {Marlon}\ \bibnamefont
  {Brenes}}, \bibinfo {author} {\bibfnamefont {Tyler}\ \bibnamefont {LeBlond}},
  \bibinfo {author} {\bibfnamefont {John}\ \bibnamefont {Goold}}, \ and\
  \bibinfo {author} {\bibfnamefont {Marcos}\ \bibnamefont {Rigol}},\ }\bibfield
   {title} {\enquote {\bibinfo {title} {Eigenstate {Thermalization} in a
  {Locally} {Perturbed} {Integrable} {System}},}\ }\href
  {https://link.aps.org/doi/10.1103/PhysRevLett.125.070605} {\bibfield
  {journal} {\bibinfo  {journal} {Phys. Rev. Lett.}\ }\textbf {\bibinfo
  {volume} {125}},\ \bibinfo {pages} {070605} (\bibinfo {year}
  {2020})}\BibitemShut {NoStop}%
\bibitem [{\citenamefont {Rigol}(2016)}]{rigol_fundamental_2016}%
  \BibitemOpen
  \bibfield  {author} {\bibinfo {author} {\bibfnamefont {Marcos}\ \bibnamefont
  {Rigol}},\ }\bibfield  {title} {\enquote {\bibinfo {title} {Fundamental
  {Asymmetry} in {Quenches} {Between} {Integrable} and {Nonintegrable}
  {Systems}},}\ }\href
  {https://link.aps.org/doi/10.1103/PhysRevLett.116.100601} {\bibfield
  {journal} {\bibinfo  {journal} {Phys. Rev. Lett.}\ }\textbf {\bibinfo
  {volume} {116}},\ \bibinfo {pages} {100601} (\bibinfo {year}
  {2016})}\BibitemShut {NoStop}%
\bibitem [{\citenamefont {Essler}\ and\ \citenamefont
  {Fagotti}(2016)}]{essler_quench_2016}%
  \BibitemOpen
  \bibfield  {author} {\bibinfo {author} {\bibfnamefont {Fabian H.~L.}\
  \bibnamefont {Essler}}\ and\ \bibinfo {author} {\bibfnamefont {Maurizio}\
  \bibnamefont {Fagotti}},\ }\bibfield  {title} {\enquote {\bibinfo {title}
  {Quench dynamics and relaxation in isolated integrable quantum spin
  chains},}\ }\href {https://dx.doi.org/10.1088/1742-5468/2016/06/064002}
  {\bibfield  {journal} {\bibinfo  {journal} {J. Stat. Mech.}\ }\textbf
  {\bibinfo {volume} {2016}},\ \bibinfo {pages} {064002} (\bibinfo {year}
  {2016})}\BibitemShut {NoStop}%
\bibitem [{\citenamefont {Rigol}\ \emph {et~al.}(2007)\citenamefont {Rigol},
  \citenamefont {Dunjko}, \citenamefont {Yurovsky},\ and\ \citenamefont
  {Olshanii}}]{rigol_relaxation_2007}%
  \BibitemOpen
  \bibfield  {author} {\bibinfo {author} {\bibfnamefont {Marcos}\ \bibnamefont
  {Rigol}}, \bibinfo {author} {\bibfnamefont {Vanja}\ \bibnamefont {Dunjko}},
  \bibinfo {author} {\bibfnamefont {Vladimir}\ \bibnamefont {Yurovsky}}, \ and\
  \bibinfo {author} {\bibfnamefont {Maxim}\ \bibnamefont {Olshanii}},\
  }\bibfield  {title} {\enquote {\bibinfo {title} {Relaxation in a {Completely}
  {Integrable} {Many}-{Body} {Quantum} {System}: {An} {Ab} {Initio} {Study} of
  the {Dynamics} of the {Highly} {Excited} {States} of {1D} {Lattice}
  {Hard}-{Core} {Bosons}},}\ }\href
  {https://link.aps.org/doi/10.1103/PhysRevLett.98.050405} {\bibfield
  {journal} {\bibinfo  {journal} {Phys. Rev. Lett.}\ }\textbf {\bibinfo
  {volume} {98}},\ \bibinfo {pages} {050405} (\bibinfo {year}
  {2007})}\BibitemShut {NoStop}%
\bibitem [{\citenamefont {Basko}\ \emph {et~al.}(2006)\citenamefont {Basko},
  \citenamefont {Aleiner},\ and\ \citenamefont
  {Altshuler}}]{basko_metalinsulator_2006}%
  \BibitemOpen
  \bibfield  {author} {\bibinfo {author} {\bibfnamefont {D.~M.}\ \bibnamefont
  {Basko}}, \bibinfo {author} {\bibfnamefont {I.~L.}\ \bibnamefont {Aleiner}},
  \ and\ \bibinfo {author} {\bibfnamefont {B.~L.}\ \bibnamefont {Altshuler}},\
  }\bibfield  {title} {\enquote {\bibinfo {title} {Metal–insulator transition
  in a weakly interacting many-electron system with localized single-particle
  states},}\ }\href
  {https://www.sciencedirect.com/science/article/pii/S0003491605002630}
  {\bibfield  {journal} {\bibinfo  {journal} {Annals of Physics}\ }\textbf
  {\bibinfo {volume} {321}},\ \bibinfo {pages} {1126--1205} (\bibinfo {year}
  {2006})}\BibitemShut {NoStop}%
\bibitem [{\citenamefont {Gornyi}\ \emph {et~al.}(2005)\citenamefont {Gornyi},
  \citenamefont {Mirlin},\ and\ \citenamefont
  {Polyakov}}]{gornyi_interacting_2005}%
  \BibitemOpen
  \bibfield  {author} {\bibinfo {author} {\bibfnamefont {I.~V.}\ \bibnamefont
  {Gornyi}}, \bibinfo {author} {\bibfnamefont {A.~D.}\ \bibnamefont {Mirlin}},
  \ and\ \bibinfo {author} {\bibfnamefont {D.~G.}\ \bibnamefont {Polyakov}},\
  }\bibfield  {title} {\enquote {\bibinfo {title} {Interacting {Electrons} in
  {Disordered} {Wires}: {Anderson} {Localization} and {Low}-\${T}\$
  {Transport}},}\ }\href
  {https://link.aps.org/doi/10.1103/PhysRevLett.95.206603} {\bibfield
  {journal} {\bibinfo  {journal} {Phys. Rev. Lett.}\ }\textbf {\bibinfo
  {volume} {95}},\ \bibinfo {pages} {206603} (\bibinfo {year}
  {2005})}\BibitemShut {NoStop}%
\bibitem [{\citenamefont {Oganesyan}\ and\ \citenamefont
  {Huse}(2007)}]{oganesyan_localization_2007}%
  \BibitemOpen
  \bibfield  {author} {\bibinfo {author} {\bibfnamefont {Vadim}\ \bibnamefont
  {Oganesyan}}\ and\ \bibinfo {author} {\bibfnamefont {David~A.}\ \bibnamefont
  {Huse}},\ }\bibfield  {title} {\enquote {\bibinfo {title} {Localization of
  interacting fermions at high temperature},}\ }\href
  {https://link.aps.org/doi/10.1103/PhysRevB.75.155111} {\bibfield  {journal}
  {\bibinfo  {journal} {Phys. Rev. B}\ }\textbf {\bibinfo {volume} {75}},\
  \bibinfo {pages} {155111} (\bibinfo {year} {2007})}\BibitemShut {NoStop}%
\bibitem [{\citenamefont {Huse}\ \emph {et~al.}(2014)\citenamefont {Huse},
  \citenamefont {Nandkishore},\ and\ \citenamefont
  {Oganesyan}}]{huse_phenomenology_2014}%
  \BibitemOpen
  \bibfield  {author} {\bibinfo {author} {\bibfnamefont {David~A.}\
  \bibnamefont {Huse}}, \bibinfo {author} {\bibfnamefont {Rahul}\ \bibnamefont
  {Nandkishore}}, \ and\ \bibinfo {author} {\bibfnamefont {Vadim}\ \bibnamefont
  {Oganesyan}},\ }\bibfield  {title} {\enquote {\bibinfo {title} {Phenomenology
  of fully many-body-localized systems},}\ }\href
  {https://link.aps.org/doi/10.1103/PhysRevB.90.174202} {\bibfield  {journal}
  {\bibinfo  {journal} {Phys. Rev. B}\ }\textbf {\bibinfo {volume} {90}},\
  \bibinfo {pages} {174202} (\bibinfo {year} {2014})}\BibitemShut {NoStop}%
\bibitem [{\citenamefont {Serbyn}\ \emph {et~al.}(2013)\citenamefont {Serbyn},
  \citenamefont {Papic},\ and\ \citenamefont {Abanin}}]{serbyn_local_2013}%
  \BibitemOpen
  \bibfield  {author} {\bibinfo {author} {\bibfnamefont {Maksym}\ \bibnamefont
  {Serbyn}}, \bibinfo {author} {\bibfnamefont {Z.}~\bibnamefont {Papic}}, \
  and\ \bibinfo {author} {\bibfnamefont {Dmitry~A.}\ \bibnamefont {Abanin}},\
  }\bibfield  {title} {\enquote {\bibinfo {title} {Local {Conservation} {Laws}
  and the {Structure} of the {Many}-{Body} {Localized} {States}},}\ }\href
  {https://link.aps.org/doi/10.1103/PhysRevLett.111.127201} {\bibfield
  {journal} {\bibinfo  {journal} {Phys. Rev. Lett.}\ }\textbf {\bibinfo
  {volume} {111}},\ \bibinfo {pages} {127201} (\bibinfo {year}
  {2013})}\BibitemShut {NoStop}%
\bibitem [{\citenamefont {Schreiber}\ \emph {et~al.}(2015)\citenamefont
  {Schreiber}, \citenamefont {Hodgman}, \citenamefont {Bordia}, \citenamefont
  {Lüschen}, \citenamefont {Fischer}, \citenamefont {Vosk}, \citenamefont
  {Altman}, \citenamefont {Schneider},\ and\ \citenamefont
  {Bloch}}]{schreiber_observation_2015}%
  \BibitemOpen
  \bibfield  {author} {\bibinfo {author} {\bibfnamefont {Michael}\ \bibnamefont
  {Schreiber}}, \bibinfo {author} {\bibfnamefont {Sean~S.}\ \bibnamefont
  {Hodgman}}, \bibinfo {author} {\bibfnamefont {Pranjal}\ \bibnamefont
  {Bordia}}, \bibinfo {author} {\bibfnamefont {Henrik~P.}\ \bibnamefont
  {Lüschen}}, \bibinfo {author} {\bibfnamefont {Mark~H.}\ \bibnamefont
  {Fischer}}, \bibinfo {author} {\bibfnamefont {Ronen}\ \bibnamefont {Vosk}},
  \bibinfo {author} {\bibfnamefont {Ehud}\ \bibnamefont {Altman}}, \bibinfo
  {author} {\bibfnamefont {Ulrich}\ \bibnamefont {Schneider}}, \ and\ \bibinfo
  {author} {\bibfnamefont {Immanuel}\ \bibnamefont {Bloch}},\ }\bibfield
  {title} {\enquote {\bibinfo {title} {Observation of many-body localization of
  interacting fermions in a quasirandom optical lattice},}\ }\href
  {https://www.science.org/doi/10.1126/science.aaa7432} {\bibfield  {journal}
  {\bibinfo  {journal} {Science}\ }\textbf {\bibinfo {volume} {349}},\ \bibinfo
  {pages} {842--845} (\bibinfo {year} {2015})}\BibitemShut {NoStop}%
\bibitem [{\citenamefont {Luitz}\ \emph {et~al.}(2015)\citenamefont {Luitz},
  \citenamefont {Laflorencie},\ and\ \citenamefont
  {Alet}}]{luitz_many-body_2015}%
  \BibitemOpen
  \bibfield  {author} {\bibinfo {author} {\bibfnamefont {David~J.}\
  \bibnamefont {Luitz}}, \bibinfo {author} {\bibfnamefont {Nicolas}\
  \bibnamefont {Laflorencie}}, \ and\ \bibinfo {author} {\bibfnamefont
  {Fabien}\ \bibnamefont {Alet}},\ }\bibfield  {title} {\enquote {\bibinfo
  {title} {Many-body localization edge in the random-field {Heisenberg}
  chain},}\ }\href {https://link.aps.org/doi/10.1103/PhysRevB.91.081103}
  {\bibfield  {journal} {\bibinfo  {journal} {Phys. Rev. B}\ }\textbf {\bibinfo
  {volume} {91}},\ \bibinfo {pages} {081103(R)} (\bibinfo {year}
  {2015})}\BibitemShut {NoStop}%
\bibitem [{\citenamefont {Suntajs}\ \emph {et~al.}(2020)\citenamefont
  {Suntajs}, \citenamefont {Bonca}, \citenamefont {Prosen},\ and\ \citenamefont
  {Vidmar}}]{suntajs_quantum_2020}%
  \BibitemOpen
  \bibfield  {author} {\bibinfo {author} {\bibfnamefont {Jan}\ \bibnamefont
  {Suntajs}}, \bibinfo {author} {\bibfnamefont {Janez}\ \bibnamefont {Bonca}},
  \bibinfo {author} {\bibfnamefont {Tomaž}\ \bibnamefont {Prosen}}, \ and\
  \bibinfo {author} {\bibfnamefont {Lev}\ \bibnamefont {Vidmar}},\ }\bibfield
  {title} {\enquote {\bibinfo {title} {Quantum chaos challenges many-body
  localization},}\ }\href
  {https://link.aps.org/doi/10.1103/PhysRevE.102.062144} {\bibfield  {journal}
  {\bibinfo  {journal} {Phys. Rev. E}\ }\textbf {\bibinfo {volume} {102}},\
  \bibinfo {pages} {062144} (\bibinfo {year} {2020})}\BibitemShut {NoStop}%
\bibitem [{\citenamefont {Sels}\ and\ \citenamefont
  {Polkovnikov}(2023)}]{sels_thermalization_2023}%
  \BibitemOpen
  \bibfield  {author} {\bibinfo {author} {\bibfnamefont {Dries}\ \bibnamefont
  {Sels}}\ and\ \bibinfo {author} {\bibfnamefont {Anatoli}\ \bibnamefont
  {Polkovnikov}},\ }\bibfield  {title} {\enquote {\bibinfo {title}
  {Thermalization of {Dilute} {Impurities} in {One}-{Dimensional} {Spin}
  {Chains}},}\ }\href {https://link.aps.org/doi/10.1103/PhysRevX.13.011041}
  {\bibfield  {journal} {\bibinfo  {journal} {Phys. Rev. X}\ }\textbf {\bibinfo
  {volume} {13}},\ \bibinfo {pages} {011041} (\bibinfo {year}
  {2023})}\BibitemShut {NoStop}%
\bibitem [{\citenamefont {Sels}\ and\ \citenamefont
  {Polkovnikov}(2021)}]{sels_dynamical_2021}%
  \BibitemOpen
  \bibfield  {author} {\bibinfo {author} {\bibfnamefont {Dries}\ \bibnamefont
  {Sels}}\ and\ \bibinfo {author} {\bibfnamefont {Anatoli}\ \bibnamefont
  {Polkovnikov}},\ }\bibfield  {title} {\enquote {\bibinfo {title} {Dynamical
  obstruction to localization in a disordered spin chain},}\ }\href
  {https://link.aps.org/doi/10.1103/PhysRevE.104.054105} {\bibfield  {journal}
  {\bibinfo  {journal} {Phys. Rev. E}\ }\textbf {\bibinfo {volume} {104}},\
  \bibinfo {pages} {054105} (\bibinfo {year} {2021})}\BibitemShut {NoStop}%
\bibitem [{\citenamefont {Kiefer-Emmanouilidis}\ \emph {et~al.}()\citenamefont
  {Kiefer-Emmanouilidis}, \citenamefont {Unanyan}, \citenamefont
  {Fleischhauer},\ and\ \citenamefont
  {Sirker}}]{kiefer-emmanouilidis_slow_2021}%
  \BibitemOpen
  \bibfield  {author} {\bibinfo {author} {\bibfnamefont {Maximilian}\
  \bibnamefont {Kiefer-Emmanouilidis}}, \bibinfo {author} {\bibfnamefont
  {Razmik}\ \bibnamefont {Unanyan}}, \bibinfo {author} {\bibfnamefont
  {Michael}\ \bibnamefont {Fleischhauer}}, \ and\ \bibinfo {author}
  {\bibfnamefont {Jesko}\ \bibnamefont {Sirker}},\ }\bibfield  {title}
  {\enquote {\bibinfo {title} {Slow delocalization of particles in many-body
  localized phases},}\ }\href
  {https://link.aps.org/doi/10.1103/PhysRevB.103.024203} {\bibfield  {journal}
  {\bibinfo  {journal} {Phys. Rev. B}\ }\textbf {\bibinfo {volume} {103}},\
  \bibinfo {pages} {024203}}\BibitemShut {NoStop}%
\bibitem [{\citenamefont {Morningstar}\ \emph {et~al.}(2022)\citenamefont
  {Morningstar}, \citenamefont {Colmenarez}, \citenamefont {Khemani},
  \citenamefont {Luitz},\ and\ \citenamefont
  {Huse}}]{morningstar_avalanches_2022}%
  \BibitemOpen
  \bibfield  {author} {\bibinfo {author} {\bibfnamefont {Alan}\ \bibnamefont
  {Morningstar}}, \bibinfo {author} {\bibfnamefont {Luis}\ \bibnamefont
  {Colmenarez}}, \bibinfo {author} {\bibfnamefont {Vedika}\ \bibnamefont
  {Khemani}}, \bibinfo {author} {\bibfnamefont {David~J.}\ \bibnamefont
  {Luitz}}, \ and\ \bibinfo {author} {\bibfnamefont {David~A.}\ \bibnamefont
  {Huse}},\ }\bibfield  {title} {\enquote {\bibinfo {title} {Avalanches and
  many-body resonances in many-body localized systems},}\ }\href
  {https://link.aps.org/doi/10.1103/PhysRevB.105.174205} {\bibfield  {journal}
  {\bibinfo  {journal} {Phys. Rev. B}\ }\textbf {\bibinfo {volume} {105}},\
  \bibinfo {pages} {174205} (\bibinfo {year} {2022})}\BibitemShut {NoStop}%
\bibitem [{\citenamefont {Sels}(2022)}]{sels_bath-induced_2022}%
  \BibitemOpen
  \bibfield  {author} {\bibinfo {author} {\bibfnamefont {Dries}\ \bibnamefont
  {Sels}},\ }\bibfield  {title} {\enquote {\bibinfo {title} {Bath-induced
  delocalization in interacting disordered spin chains},}\ }\href
  {https://link.aps.org/doi/10.1103/PhysRevB.106.L020202} {\bibfield  {journal}
  {\bibinfo  {journal} {Phys. Rev. B}\ }\textbf {\bibinfo {volume} {106}},\
  \bibinfo {pages} {L020202} (\bibinfo {year} {2022})}\BibitemShut {NoStop}%
\bibitem [{\citenamefont {Crowley}\ and\ \citenamefont
  {Chandran}(2020)}]{crowley_avalanche_2020}%
  \BibitemOpen
  \bibfield  {author} {\bibinfo {author} {\bibfnamefont {P.~J.~D.}\
  \bibnamefont {Crowley}}\ and\ \bibinfo {author} {\bibfnamefont
  {A.}~\bibnamefont {Chandran}},\ }\bibfield  {title} {\enquote {\bibinfo
  {title} {Avalanche induced coexisting localized and thermal regions in
  disordered chains},}\ }\href
  {https://link.aps.org/doi/10.1103/PhysRevResearch.2.033262} {\bibfield
  {journal} {\bibinfo  {journal} {Phys. Rev. Res.}\ }\textbf {\bibinfo {volume}
  {2}},\ \bibinfo {pages} {033262} (\bibinfo {year} {2020})}\BibitemShut
  {NoStop}%
\bibitem [{\citenamefont {Sierant}\ and\ \citenamefont
  {Zakrzewski}(2022)}]{sierant_challenges_2022}%
  \BibitemOpen
  \bibfield  {author} {\bibinfo {author} {\bibfnamefont {Piotr}\ \bibnamefont
  {Sierant}}\ and\ \bibinfo {author} {\bibfnamefont {Jakub}\ \bibnamefont
  {Zakrzewski}},\ }\bibfield  {title} {\enquote {\bibinfo {title} {Challenges
  to observation of many-body localization},}\ }\href
  {https://link.aps.org/doi/10.1103/PhysRevB.105.224203} {\bibfield  {journal}
  {\bibinfo  {journal} {Phys. Rev. B}\ }\textbf {\bibinfo {volume} {105}},\
  \bibinfo {pages} {224203} (\bibinfo {year} {2022})}\BibitemShut {NoStop}%
\bibitem [{\citenamefont {Sierant}\ \emph {et~al.}()\citenamefont {Sierant},
  \citenamefont {Delande},\ and\ \citenamefont
  {Zakrzewski}}]{sierant_thouless_2020}%
  \BibitemOpen
  \bibfield  {author} {\bibinfo {author} {\bibfnamefont {Piotr}\ \bibnamefont
  {Sierant}}, \bibinfo {author} {\bibfnamefont {Dominique}\ \bibnamefont
  {Delande}}, \ and\ \bibinfo {author} {\bibfnamefont {Jakub}\ \bibnamefont
  {Zakrzewski}},\ }\bibfield  {title} {\enquote {\bibinfo {title} {Thouless
  {Time} {Analysis} of {Anderson} and {Many}-{Body} {Localization}
  {Transitions}},}\ }\href
  {https://link.aps.org/doi/10.1103/PhysRevLett.124.186601} {\bibfield
  {journal} {\bibinfo  {journal} {Phys. Rev. Lett.}\ }\textbf {\bibinfo
  {volume} {124}},\ \bibinfo {pages} {186601}}\BibitemShut {NoStop}%
\bibitem [{\citenamefont {Panda}\ \emph {et~al.}(2020)\citenamefont {Panda},
  \citenamefont {Scardicchio}, \citenamefont {Schulz}, \citenamefont {Taylor},\
  and\ \citenamefont {Žnidarič}}]{panda_can_2020}%
  \BibitemOpen
  \bibfield  {author} {\bibinfo {author} {\bibfnamefont {R.~K.}\ \bibnamefont
  {Panda}}, \bibinfo {author} {\bibfnamefont {A.}~\bibnamefont {Scardicchio}},
  \bibinfo {author} {\bibfnamefont {M.}~\bibnamefont {Schulz}}, \bibinfo
  {author} {\bibfnamefont {S.~R.}\ \bibnamefont {Taylor}}, \ and\ \bibinfo
  {author} {\bibfnamefont {M.}~\bibnamefont {Žnidarič}},\ }\bibfield  {title}
  {\enquote {\bibinfo {title} {Can we study the many-body localisation
  transition?}}\ }\href {https://dx.doi.org/10.1209/0295-5075/128/67003}
  {\bibfield  {journal} {\bibinfo  {journal} {EPL}\ }\textbf {\bibinfo {volume}
  {128}},\ \bibinfo {pages} {67003} (\bibinfo {year} {2020})}\BibitemShut
  {NoStop}%
\bibitem [{\citenamefont {Crowley}\ and\ \citenamefont
  {Chandran}(2022)}]{crowley_constructive_2022}%
  \BibitemOpen
  \bibfield  {author} {\bibinfo {author} {\bibfnamefont {Philip}\ \bibnamefont
  {Crowley}}\ and\ \bibinfo {author} {\bibfnamefont {Anushya}\ \bibnamefont
  {Chandran}},\ }\bibfield  {title} {\enquote {\bibinfo {title} {A constructive
  theory of the numerically accessible many-body localized to thermal
  crossover},}\ }\href
  {https://www.scipost.org/SciPostPhys.12.6.201?acad_field_slug=physics}
  {\bibfield  {journal} {\bibinfo  {journal} {SciPost Physics}\ }\textbf
  {\bibinfo {volume} {12}},\ \bibinfo {pages} {201} (\bibinfo {year}
  {2022})}\BibitemShut {NoStop}%
\bibitem [{\citenamefont {Laflorencie}\ \emph {et~al.}(2022)\citenamefont
  {Laflorencie}, \citenamefont {Lemarie},\ and\ \citenamefont
  {Mace}}]{laflorencie_topological_2022}%
  \BibitemOpen
  \bibfield  {author} {\bibinfo {author} {\bibfnamefont {Nicolas}\ \bibnamefont
  {Laflorencie}}, \bibinfo {author} {\bibfnamefont {Gabriel}\ \bibnamefont
  {Lemarie}}, \ and\ \bibinfo {author} {\bibfnamefont {Nicolas}\ \bibnamefont
  {Mace}},\ }\bibfield  {title} {\enquote {\bibinfo {title} {Topological order
  in random interacting {Ising}-{Majorana} chains stabilized by many-body
  localization},}\ }\href
  {https://link.aps.org/doi/10.1103/PhysRevResearch.4.L032016} {\bibfield
  {journal} {\bibinfo  {journal} {Phys. Rev. Res.}\ }\textbf {\bibinfo {volume}
  {4}},\ \bibinfo {pages} {L032016} (\bibinfo {year} {2022})}\BibitemShut
  {NoStop}%
\bibitem [{\citenamefont {De~Roeck}\ and\ \citenamefont
  {Imbrie}(2017)}]{de_roeck_many-body_2017}%
  \BibitemOpen
  \bibfield  {author} {\bibinfo {author} {\bibfnamefont {Wojciech}\
  \bibnamefont {De~Roeck}}\ and\ \bibinfo {author} {\bibfnamefont {John~Z.}\
  \bibnamefont {Imbrie}},\ }\bibfield  {title} {\enquote {\bibinfo {title}
  {Many-body localization: stability and instability},}\ }\href
  {https://royalsocietypublishing.org/doi/10.1098/rsta.2016.0422} {\bibfield
  {journal} {\bibinfo  {journal} {Philosophical Transactions of the Royal
  Society A: Mathematical, Physical and Engineering Sciences}\ }\textbf
  {\bibinfo {volume} {375}},\ \bibinfo {pages} {20160422} (\bibinfo {year}
  {2017})}\BibitemShut {NoStop}%
\bibitem [{\citenamefont {Thiery}\ \emph {et~al.}(2018)\citenamefont {Thiery},
  \citenamefont {Huveneers}, \citenamefont {Muller},\ and\ \citenamefont
  {De~Roeck}}]{thiery_many-body_2018}%
  \BibitemOpen
  \bibfield  {author} {\bibinfo {author} {\bibfnamefont {Thimothée}\
  \bibnamefont {Thiery}}, \bibinfo {author} {\bibfnamefont {François}\
  \bibnamefont {Huveneers}}, \bibinfo {author} {\bibfnamefont {Markus}\
  \bibnamefont {Muller}}, \ and\ \bibinfo {author} {\bibfnamefont {Wojciech}\
  \bibnamefont {De~Roeck}},\ }\bibfield  {title} {\enquote {\bibinfo {title}
  {Many-{Body} {Delocalization} as a {Quantum} {Avalanche}},}\ }\href
  {https://link.aps.org/doi/10.1103/PhysRevLett.121.140601} {\bibfield
  {journal} {\bibinfo  {journal} {Phys. Rev. Lett.}\ }\textbf {\bibinfo
  {volume} {121}},\ \bibinfo {pages} {140601} (\bibinfo {year}
  {2018})}\BibitemShut {NoStop}%
\bibitem [{\citenamefont {Krajewski}\ \emph {et~al.}(2022)\citenamefont
  {Krajewski}, \citenamefont {Vidmar}, \citenamefont {Bonca},\ and\
  \citenamefont {Mierzejewski}}]{krajewski_restoring_2022}%
  \BibitemOpen
  \bibfield  {author} {\bibinfo {author} {\bibfnamefont {B.}~\bibnamefont
  {Krajewski}}, \bibinfo {author} {\bibfnamefont {L.}~\bibnamefont {Vidmar}},
  \bibinfo {author} {\bibfnamefont {J.}~\bibnamefont {Bonca}}, \ and\ \bibinfo
  {author} {\bibfnamefont {M.}~\bibnamefont {Mierzejewski}},\ }\bibfield
  {title} {\enquote {\bibinfo {title} {Restoring {Ergodicity} in a {Strongly}
  {Disordered} {Interacting} {Chain}},}\ }\href
  {https://link.aps.org/doi/10.1103/PhysRevLett.129.260601} {\bibfield
  {journal} {\bibinfo  {journal} {Phys. Rev. Lett.}\ }\textbf {\bibinfo
  {volume} {129}},\ \bibinfo {pages} {260601} (\bibinfo {year}
  {2022})}\BibitemShut {NoStop}%
\bibitem [{\citenamefont {Bardarson}\ \emph {et~al.}(2012)\citenamefont
  {Bardarson}, \citenamefont {Pollmann},\ and\ \citenamefont
  {Moore}}]{bardarson_unbounded_2012}%
  \BibitemOpen
  \bibfield  {author} {\bibinfo {author} {\bibfnamefont {Jens~H.}\ \bibnamefont
  {Bardarson}}, \bibinfo {author} {\bibfnamefont {Frank}\ \bibnamefont
  {Pollmann}}, \ and\ \bibinfo {author} {\bibfnamefont {Joel~E.}\ \bibnamefont
  {Moore}},\ }\bibfield  {title} {\enquote {\bibinfo {title} {Unbounded
  {Growth} of {Entanglement} in {Models} of {Many}-{Body} {Localization}},}\
  }\href {https://link.aps.org/doi/10.1103/PhysRevLett.109.017202} {\bibfield
  {journal} {\bibinfo  {journal} {Phys. Rev. Lett.}\ }\textbf {\bibinfo
  {volume} {109}},\ \bibinfo {pages} {017202} (\bibinfo {year}
  {2012})}\BibitemShut {NoStop}%
\bibitem [{\citenamefont {Serbyn}\ and\ \citenamefont
  {Moore}(2016)}]{serbyn_spectral_2016}%
  \BibitemOpen
  \bibfield  {author} {\bibinfo {author} {\bibfnamefont {Maksym}\ \bibnamefont
  {Serbyn}}\ and\ \bibinfo {author} {\bibfnamefont {Joel~E.}\ \bibnamefont
  {Moore}},\ }\bibfield  {title} {\enquote {\bibinfo {title} {Spectral
  statistics across the many-body localization transition},}\ }\href {\doibase
  10.1103/PhysRevB.93.041424} {\bibfield  {journal} {\bibinfo  {journal} {Phys.
  Rev. B}\ }\textbf {\bibinfo {volume} {93}},\ \bibinfo {pages} {041424(R)}
  (\bibinfo {year} {2016})},\ \bibinfo {note} {publisher: American Physical
  Society}\BibitemShut {NoStop}%
\bibitem [{\citenamefont {Ray}\ \emph {et~al.}(2018)\citenamefont {Ray},
  \citenamefont {Ghosh},\ and\ \citenamefont {Sinha}}]{ray_drive-induced_2018}%
  \BibitemOpen
  \bibfield  {author} {\bibinfo {author} {\bibfnamefont {S.}~\bibnamefont
  {Ray}}, \bibinfo {author} {\bibfnamefont {A.}~\bibnamefont {Ghosh}}, \ and\
  \bibinfo {author} {\bibfnamefont {S.}~\bibnamefont {Sinha}},\ }\bibfield
  {title} {\enquote {\bibinfo {title} {Drive-induced delocalization in the
  {Aubry}-{Andr}{\textbackslash}'e model},}\ }\href {\doibase
  10.1103/PhysRevE.97.010101} {\bibfield  {journal} {\bibinfo  {journal} {Phys.
  Rev. E}\ }\textbf {\bibinfo {volume} {97}},\ \bibinfo {pages} {010101(R)}
  (\bibinfo {year} {2018})},\ \bibinfo {note} {publisher: American Physical
  Society}\BibitemShut {NoStop}%
\bibitem [{\citenamefont {Lazarides}\ \emph {et~al.}(2015)\citenamefont
  {Lazarides}, \citenamefont {Das},\ and\ \citenamefont
  {Moessner}}]{lazarides_fate_2015}%
  \BibitemOpen
  \bibfield  {author} {\bibinfo {author} {\bibfnamefont {Achilleas}\
  \bibnamefont {Lazarides}}, \bibinfo {author} {\bibfnamefont {Arnab}\
  \bibnamefont {Das}}, \ and\ \bibinfo {author} {\bibfnamefont {Roderich}\
  \bibnamefont {Moessner}},\ }\bibfield  {title} {\enquote {\bibinfo {title}
  {Fate of {Many}-{Body} {Localization} {Under} {Periodic} {Driving}},}\ }\href
  {\doibase 10.1103/PhysRevLett.115.030402} {\bibfield  {journal} {\bibinfo
  {journal} {Phys. Rev. Lett.}\ }\textbf {\bibinfo {volume} {115}},\ \bibinfo
  {pages} {030402} (\bibinfo {year} {2015})},\ \bibinfo {note} {publisher:
  American Physical Society}\BibitemShut {NoStop}%
\bibitem [{\citenamefont {Znidaric}\ \emph {et~al.}(2008)\citenamefont
  {Znidaric}, \citenamefont {Prosen},\ and\ \citenamefont
  {Prelovsek}}]{znidaric_many-body_2008}%
  \BibitemOpen
  \bibfield  {author} {\bibinfo {author} {\bibfnamefont {Marko}\ \bibnamefont
  {Znidaric}}, \bibinfo {author} {\bibfnamefont {Tomaz}\ \bibnamefont
  {Prosen}}, \ and\ \bibinfo {author} {\bibfnamefont {Peter}\ \bibnamefont
  {Prelovsek}},\ }\bibfield  {title} {\enquote {\bibinfo {title} {Many-body
  localization in the {Heisenberg} \${XXZ}\$ magnet in a random field},}\
  }\href {https://link.aps.org/doi/10.1103/PhysRevB.77.064426} {\bibfield
  {journal} {\bibinfo  {journal} {Phys. Rev. B}\ }\textbf {\bibinfo {volume}
  {77}},\ \bibinfo {pages} {064426} (\bibinfo {year} {2008})}\BibitemShut
  {NoStop}%
\bibitem [{\citenamefont {Sierant}\ \emph {et~al.}(2023)\citenamefont
  {Sierant}, \citenamefont {Lewenstein}, \citenamefont {Scardicchio},\ and\
  \citenamefont {Zakrzewski}}]{sierant_stability_2023}%
  \BibitemOpen
  \bibfield  {author} {\bibinfo {author} {\bibfnamefont {Piotr}\ \bibnamefont
  {Sierant}}, \bibinfo {author} {\bibfnamefont {Maciej}\ \bibnamefont
  {Lewenstein}}, \bibinfo {author} {\bibfnamefont {Antonello}\ \bibnamefont
  {Scardicchio}}, \ and\ \bibinfo {author} {\bibfnamefont {Jakub}\ \bibnamefont
  {Zakrzewski}},\ }\bibfield  {title} {\enquote {\bibinfo {title} {Stability of
  many-body localization in {Floquet} systems},}\ }\href {\doibase
  10.1103/PhysRevB.107.115132} {\bibfield  {journal} {\bibinfo  {journal}
  {Phys. Rev. B}\ }\textbf {\bibinfo {volume} {107}},\ \bibinfo {pages}
  {115132} (\bibinfo {year} {2023})},\ \bibinfo {note} {publisher: American
  Physical Society}\BibitemShut {NoStop}%
\bibitem [{\citenamefont {Bulchandani}\ \emph {et~al.}()\citenamefont
  {Bulchandani}, \citenamefont {Huse},\ and\ \citenamefont
  {Gopalakrishnan}}]{bulchandani_onset_2022}%
  \BibitemOpen
  \bibfield  {author} {\bibinfo {author} {\bibfnamefont {Vir~B.}\ \bibnamefont
  {Bulchandani}}, \bibinfo {author} {\bibfnamefont {David~A.}\ \bibnamefont
  {Huse}}, \ and\ \bibinfo {author} {\bibfnamefont {Sarang}\ \bibnamefont
  {Gopalakrishnan}},\ }\bibfield  {title} {\enquote {\bibinfo {title} {Onset of
  many-body quantum chaos due to breaking integrability},}\ }\href
  {https://link.aps.org/doi/10.1103/PhysRevB.105.214308} {\bibfield  {journal}
  {\bibinfo  {journal} {Phys. Rev. B}\ }\textbf {\bibinfo {volume} {105}},\
  \bibinfo {pages} {214308}}\BibitemShut {NoStop}%
\bibitem [{\citenamefont {Kraus}\ and\ \citenamefont
  {Cirac}(2001)}]{kraus_optimal_2001}%
  \BibitemOpen
  \bibfield  {author} {\bibinfo {author} {\bibfnamefont {B.}~\bibnamefont
  {Kraus}}\ and\ \bibinfo {author} {\bibfnamefont {J.~I.}\ \bibnamefont
  {Cirac}},\ }\bibfield  {title} {\enquote {\bibinfo {title} {Optimal creation
  of entanglement using a two-qubit gate},}\ }\href
  {https://link.aps.org/doi/10.1103/PhysRevA.63.062309} {\bibfield  {journal}
  {\bibinfo  {journal} {Phys. Rev. A}\ }\textbf {\bibinfo {volume} {63}},\
  \bibinfo {pages} {062309} (\bibinfo {year} {2001})}\BibitemShut {NoStop}%
\bibitem [{\citenamefont {Balakrishnan}\ and\ \citenamefont
  {Sankaranarayanan}(2011{\natexlab{a}})}]{balakrishnan_measures_2011}%
  \BibitemOpen
  \bibfield  {author} {\bibinfo {author} {\bibfnamefont {S.}~\bibnamefont
  {Balakrishnan}}\ and\ \bibinfo {author} {\bibfnamefont {R.}~\bibnamefont
  {Sankaranarayanan}},\ }\bibfield  {title} {\enquote {\bibinfo {title}
  {Measures of operator entanglement of two-qubit gates},}\ }\href
  {https://link.aps.org/doi/10.1103/PhysRevA.83.062320} {\bibfield  {journal}
  {\bibinfo  {journal} {Phys. Rev. A}\ }\textbf {\bibinfo {volume} {83}},\
  \bibinfo {pages} {062320} (\bibinfo {year} {2011}{\natexlab{a}})}\BibitemShut
  {NoStop}%
\bibitem [{\citenamefont {Balakrishnan}\ and\ \citenamefont
  {Sankaranarayanan}(2010)}]{balakrishnan_entangling_2010}%
  \BibitemOpen
  \bibfield  {author} {\bibinfo {author} {\bibfnamefont {S.}~\bibnamefont
  {Balakrishnan}}\ and\ \bibinfo {author} {\bibfnamefont {R.}~\bibnamefont
  {Sankaranarayanan}},\ }\bibfield  {title} {\enquote {\bibinfo {title}
  {Entangling power and local invariants of two-qubit gates},}\ }\href
  {https://link.aps.org/doi/10.1103/PhysRevA.82.034301} {\bibfield  {journal}
  {\bibinfo  {journal} {Phys. Rev. A}\ }\textbf {\bibinfo {volume} {82}},\
  \bibinfo {pages} {034301} (\bibinfo {year} {2010})}\BibitemShut {NoStop}%
\bibitem [{\citenamefont {Balakrishnan}\ and\ \citenamefont
  {Sankaranarayanan}(2009)}]{balakrishnan_characterizing_2009}%
  \BibitemOpen
  \bibfield  {author} {\bibinfo {author} {\bibfnamefont {S.}~\bibnamefont
  {Balakrishnan}}\ and\ \bibinfo {author} {\bibfnamefont {R.}~\bibnamefont
  {Sankaranarayanan}},\ }\bibfield  {title} {\enquote {\bibinfo {title}
  {Characterizing the geometrical edges of nonlocal two-qubit gates},}\ }\href
  {https://link.aps.org/doi/10.1103/PhysRevA.79.052339} {\bibfield  {journal}
  {\bibinfo  {journal} {Phys. Rev. A}\ }\textbf {\bibinfo {volume} {79}},\
  \bibinfo {pages} {052339} (\bibinfo {year} {2009})}\BibitemShut {NoStop}%
\bibitem [{\citenamefont {Zhang}\ \emph {et~al.}(2003)\citenamefont {Zhang},
  \citenamefont {Vala}, \citenamefont {Sastry},\ and\ \citenamefont
  {Whaley}}]{zhang_geometric_2003}%
  \BibitemOpen
  \bibfield  {author} {\bibinfo {author} {\bibfnamefont {Jun}\ \bibnamefont
  {Zhang}}, \bibinfo {author} {\bibfnamefont {Jiri}\ \bibnamefont {Vala}},
  \bibinfo {author} {\bibfnamefont {Shankar}\ \bibnamefont {Sastry}}, \ and\
  \bibinfo {author} {\bibfnamefont {K.~B.}\ \bibnamefont {Whaley}},\ }\bibfield
   {title} {\enquote {\bibinfo {title} {Geometric theory of nonlocal two-qubit
  operations},}\ }\href {https://link.aps.org/doi/10.1103/PhysRevA.67.042313}
  {\bibfield  {journal} {\bibinfo  {journal} {Phys. Rev. A}\ }\textbf {\bibinfo
  {volume} {67}},\ \bibinfo {pages} {042313} (\bibinfo {year}
  {2003})}\BibitemShut {NoStop}%
\bibitem [{\citenamefont {Makhlin}(2002)}]{makhlin_nonlocal_2002}%
  \BibitemOpen
  \bibfield  {author} {\bibinfo {author} {\bibfnamefont {Yuriy}\ \bibnamefont
  {Makhlin}},\ }\bibfield  {title} {\enquote {\bibinfo {title} {Nonlocal
  {Properties} of {Two}-{Qubit} {Gates} and {Mixed} {States}, and the
  {Optimization} of {Quantum} {Computations}},}\ }\href
  {https://doi.org/10.1023/A:1022144002391} {\bibfield  {journal} {\bibinfo
  {journal} {Quantum Information Processing}\ }\textbf {\bibinfo {volume}
  {1}},\ \bibinfo {pages} {243--252} (\bibinfo {year} {2002})}\BibitemShut
  {NoStop}%
\bibitem [{\citenamefont {Balakrishnan}\ and\ \citenamefont
  {Sankaranarayanan}(2011{\natexlab{b}})}]{balakrishnan_operator-schmidt_2011}%
  \BibitemOpen
  \bibfield  {author} {\bibinfo {author} {\bibfnamefont {S.}~\bibnamefont
  {Balakrishnan}}\ and\ \bibinfo {author} {\bibfnamefont {R.}~\bibnamefont
  {Sankaranarayanan}},\ }\bibfield  {title} {\enquote {\bibinfo {title}
  {Operator-{Schmidt} decomposition and the geometrical edges of two-qubit
  gates},}\ }\href {https://doi.org/10.1007/s11128-010-0207-9} {\bibfield
  {journal} {\bibinfo  {journal} {Quantum Inf Process}\ }\textbf {\bibinfo
  {volume} {10}},\ \bibinfo {pages} {449--461} (\bibinfo {year}
  {2011}{\natexlab{b}})}\BibitemShut {NoStop}%
\bibitem [{\citenamefont {Bertini}\ \emph
  {et~al.}(2019{\natexlab{a}})\citenamefont {Bertini}, \citenamefont {Kos},\
  and\ \citenamefont {Prosen}}]{bertini_exact_2019}%
  \BibitemOpen
  \bibfield  {author} {\bibinfo {author} {\bibfnamefont {Bruno}\ \bibnamefont
  {Bertini}}, \bibinfo {author} {\bibfnamefont {Pavel}\ \bibnamefont {Kos}}, \
  and\ \bibinfo {author} {\bibfnamefont {Tomaž}\ \bibnamefont {Prosen}},\
  }\bibfield  {title} {\enquote {\bibinfo {title} {Exact {Correlation}
  {Functions} for {Dual}-{Unitary} {Lattice} {Models} in \$1+1\$
  {Dimensions}},}\ }\href
  {https://link.aps.org/doi/10.1103/PhysRevLett.123.210601} {\bibfield
  {journal} {\bibinfo  {journal} {Phys. Rev. Lett.}\ }\textbf {\bibinfo
  {volume} {123}},\ \bibinfo {pages} {210601} (\bibinfo {year}
  {2019}{\natexlab{a}})}\BibitemShut {NoStop}%
\bibitem [{\citenamefont {Fisher}\ \emph {et~al.}(2023)\citenamefont {Fisher},
  \citenamefont {Khemani}, \citenamefont {Nahum},\ and\ \citenamefont
  {Vijay}}]{fisher_random_2023}%
  \BibitemOpen
  \bibfield  {author} {\bibinfo {author} {\bibfnamefont {Matthew~P.A.}\
  \bibnamefont {Fisher}}, \bibinfo {author} {\bibfnamefont {Vedika}\
  \bibnamefont {Khemani}}, \bibinfo {author} {\bibfnamefont {Adam}\
  \bibnamefont {Nahum}}, \ and\ \bibinfo {author} {\bibfnamefont {Sagar}\
  \bibnamefont {Vijay}},\ }\bibfield  {title} {\enquote {\bibinfo {title}
  {Random {Quantum} {Circuits}},}\ }\href
  {https://doi.org/10.1146/annurev-conmatphys-031720-030658} {\bibfield
  {journal} {\bibinfo  {journal} {Annual Review of Condensed Matter Physics}\
  }\textbf {\bibinfo {volume} {14}},\ \bibinfo {pages} {335--379} (\bibinfo
  {year} {2023})}\BibitemShut {NoStop}%
\bibitem [{\citenamefont {Piroli}\ \emph {et~al.}(2020)\citenamefont {Piroli},
  \citenamefont {Bertini}, \citenamefont {Cirac},\ and\ \citenamefont
  {Prosen}}]{piroli_exact_2020}%
  \BibitemOpen
  \bibfield  {author} {\bibinfo {author} {\bibfnamefont {Lorenzo}\ \bibnamefont
  {Piroli}}, \bibinfo {author} {\bibfnamefont {Bruno}\ \bibnamefont {Bertini}},
  \bibinfo {author} {\bibfnamefont {J.~Ignacio}\ \bibnamefont {Cirac}}, \ and\
  \bibinfo {author} {\bibfnamefont {Tomaž}\ \bibnamefont {Prosen}},\
  }\bibfield  {title} {\enquote {\bibinfo {title} {Exact dynamics in
  dual-unitary quantum circuits},}\ }\href
  {https://link.aps.org/doi/10.1103/PhysRevB.101.094304} {\bibfield  {journal}
  {\bibinfo  {journal} {Phys. Rev. B}\ }\textbf {\bibinfo {volume} {101}},\
  \bibinfo {pages} {094304} (\bibinfo {year} {2020})}\BibitemShut {NoStop}%
\bibitem [{\citenamefont {Bertini}\ \emph
  {et~al.}(2020{\natexlab{a}})\citenamefont {Bertini}, \citenamefont {Kos},\
  and\ \citenamefont {Prosen}}]{bertini_operator_2020}%
  \BibitemOpen
  \bibfield  {author} {\bibinfo {author} {\bibfnamefont {Bruno}\ \bibnamefont
  {Bertini}}, \bibinfo {author} {\bibfnamefont {Pavel}\ \bibnamefont {Kos}}, \
  and\ \bibinfo {author} {\bibfnamefont {Tomaž}\ \bibnamefont {Prosen}},\
  }\bibfield  {title} {\enquote {\bibinfo {title} {Operator {Entanglement} in
  {Local} {Quantum} {Circuits} {I}: {Chaotic} {Dual}-{Unitary} {Circuits}},}\
  }\href {https://scipost.org/10.21468/SciPostPhys.8.4.067} {\bibfield
  {journal} {\bibinfo  {journal} {SciPost Physics}\ }\textbf {\bibinfo {volume}
  {8}},\ \bibinfo {pages} {067} (\bibinfo {year}
  {2020}{\natexlab{a}})}\BibitemShut {NoStop}%
\bibitem [{\citenamefont {Bertini}\ \emph
  {et~al.}(2020{\natexlab{b}})\citenamefont {Bertini}, \citenamefont {Kos},\
  and\ \citenamefont {Prosen}}]{bertini_operator_2020-1}%
  \BibitemOpen
  \bibfield  {author} {\bibinfo {author} {\bibfnamefont {Bruno}\ \bibnamefont
  {Bertini}}, \bibinfo {author} {\bibfnamefont {Pavel}\ \bibnamefont {Kos}}, \
  and\ \bibinfo {author} {\bibfnamefont {Tomaž}\ \bibnamefont {Prosen}},\
  }\bibfield  {title} {\enquote {\bibinfo {title} {Operator {Entanglement} in
  {Local} {Quantum} {Circuits} {II}: {Solitons} in {Chains} of {Qubits}},}\
  }\href {https://scipost.org/10.21468/SciPostPhys.8.4.068} {\bibfield
  {journal} {\bibinfo  {journal} {SciPost Physics}\ }\textbf {\bibinfo {volume}
  {8}},\ \bibinfo {pages} {068} (\bibinfo {year}
  {2020}{\natexlab{b}})}\BibitemShut {NoStop}%
\bibitem [{\citenamefont {Claeys}\ and\ \citenamefont
  {Lamacraft}(2021)}]{claeys_ergodic_2021}%
  \BibitemOpen
  \bibfield  {author} {\bibinfo {author} {\bibfnamefont {Pieter~W.}\
  \bibnamefont {Claeys}}\ and\ \bibinfo {author} {\bibfnamefont {Austen}\
  \bibnamefont {Lamacraft}},\ }\bibfield  {title} {\enquote {\bibinfo {title}
  {Ergodic and {Nonergodic} {Dual}-{Unitary} {Quantum} {Circuits} with
  {Arbitrary} {Local} {Hilbert} {Space} {Dimension}},}\ }\href
  {https://link.aps.org/doi/10.1103/PhysRevLett.126.100603} {\bibfield
  {journal} {\bibinfo  {journal} {Phys. Rev. Lett.}\ }\textbf {\bibinfo
  {volume} {126}},\ \bibinfo {pages} {100603} (\bibinfo {year}
  {2021})}\BibitemShut {NoStop}%
\bibitem [{\citenamefont {Aravinda}\ \emph {et~al.}(2021)\citenamefont
  {Aravinda}, \citenamefont {Rather},\ and\ \citenamefont
  {Lakshminarayan}}]{aravinda_dual-unitary_2021}%
  \BibitemOpen
  \bibfield  {author} {\bibinfo {author} {\bibfnamefont {S.}~\bibnamefont
  {Aravinda}}, \bibinfo {author} {\bibfnamefont {Suhail~Ahmad}\ \bibnamefont
  {Rather}}, \ and\ \bibinfo {author} {\bibfnamefont {Arul}\ \bibnamefont
  {Lakshminarayan}},\ }\bibfield  {title} {\enquote {\bibinfo {title} {From
  dual-unitary to quantum {Bernoulli} circuits: {Role} of the entangling power
  in constructing a quantum ergodic hierarchy},}\ }\href
  {https://link.aps.org/doi/10.1103/PhysRevResearch.3.043034} {\bibfield
  {journal} {\bibinfo  {journal} {Phys. Rev. Res.}\ }\textbf {\bibinfo {volume}
  {3}},\ \bibinfo {pages} {043034} (\bibinfo {year} {2021})}\BibitemShut
  {NoStop}%
\bibitem [{\citenamefont {Bertini}\ \emph {et~al.}(2021)\citenamefont
  {Bertini}, \citenamefont {Kos},\ and\ \citenamefont
  {Prosen}}]{bertini_random_2021}%
  \BibitemOpen
  \bibfield  {author} {\bibinfo {author} {\bibfnamefont {Bruno}\ \bibnamefont
  {Bertini}}, \bibinfo {author} {\bibfnamefont {Pavel}\ \bibnamefont {Kos}}, \
  and\ \bibinfo {author} {\bibfnamefont {Tomaž}\ \bibnamefont {Prosen}},\
  }\bibfield  {title} {\enquote {\bibinfo {title} {Random {Matrix} {Spectral}
  {Form} {Factor} of {Dual}-{Unitary} {Quantum} {Circuits}},}\ }\href
  {https://doi.org/10.1007/s00220-021-04139-2} {\bibfield  {journal} {\bibinfo
  {journal} {Commun. Math. Phys.}\ }\textbf {\bibinfo {volume} {387}},\
  \bibinfo {pages} {597--620} (\bibinfo {year} {2021})}\BibitemShut {NoStop}%
\bibitem [{\citenamefont {Bertini}\ \emph {et~al.}(2018)\citenamefont
  {Bertini}, \citenamefont {Kos},\ and\ \citenamefont
  {Prosen}}]{bertini_exact_2018}%
  \BibitemOpen
  \bibfield  {author} {\bibinfo {author} {\bibfnamefont {Bruno}\ \bibnamefont
  {Bertini}}, \bibinfo {author} {\bibfnamefont {Pavel}\ \bibnamefont {Kos}}, \
  and\ \bibinfo {author} {\bibfnamefont {Tomaž}\ \bibnamefont {Prosen}},\
  }\bibfield  {title} {\enquote {\bibinfo {title} {Exact {Spectral} {Form}
  {Factor} in a {Minimal} {Model} of {Many}-{Body} {Quantum} {Chaos}},}\ }\href
  {https://link.aps.org/doi/10.1103/PhysRevLett.121.264101} {\bibfield
  {journal} {\bibinfo  {journal} {Phys. Rev. Lett.}\ }\textbf {\bibinfo
  {volume} {121}},\ \bibinfo {pages} {264101} (\bibinfo {year}
  {2018})}\BibitemShut {NoStop}%
\bibitem [{\citenamefont {Borsi}\ and\ \citenamefont
  {Pozsgay}(2022)}]{borsi_construction_2022}%
  \BibitemOpen
  \bibfield  {author} {\bibinfo {author} {\bibfnamefont {Márton}\ \bibnamefont
  {Borsi}}\ and\ \bibinfo {author} {\bibfnamefont {Balázs}\ \bibnamefont
  {Pozsgay}},\ }\bibfield  {title} {\enquote {\bibinfo {title} {Construction
  and the ergodicity properties of dual unitary quantum circuits},}\ }\href
  {https://link.aps.org/doi/10.1103/PhysRevB.106.014302} {\bibfield  {journal}
  {\bibinfo  {journal} {Phys. Rev. B}\ }\textbf {\bibinfo {volume} {106}},\
  \bibinfo {pages} {014302} (\bibinfo {year} {2022})}\BibitemShut {NoStop}%
\bibitem [{\citenamefont {Bertini}\ \emph
  {et~al.}(2019{\natexlab{b}})\citenamefont {Bertini}, \citenamefont {Kos},\
  and\ \citenamefont {Prosen}}]{bertini_entanglement_2019}%
  \BibitemOpen
  \bibfield  {author} {\bibinfo {author} {\bibfnamefont {Bruno}\ \bibnamefont
  {Bertini}}, \bibinfo {author} {\bibfnamefont {Pavel}\ \bibnamefont {Kos}}, \
  and\ \bibinfo {author} {\bibfnamefont {Tomaž}\ \bibnamefont {Prosen}},\
  }\bibfield  {title} {\enquote {\bibinfo {title} {Entanglement {Spreading} in
  a {Minimal} {Model} of {Maximal} {Many}-{Body} {Quantum} {Chaos}},}\ }\href
  {https://link.aps.org/doi/10.1103/PhysRevX.9.021033} {\bibfield  {journal}
  {\bibinfo  {journal} {Phys. Rev. X}\ }\textbf {\bibinfo {volume} {9}},\
  \bibinfo {pages} {021033} (\bibinfo {year} {2019}{\natexlab{b}})}\BibitemShut
  {NoStop}%
\bibitem [{\citenamefont {Bertini}\ and\ \citenamefont
  {Piroli}(2020)}]{bertini_scrambling_2020}%
  \BibitemOpen
  \bibfield  {author} {\bibinfo {author} {\bibfnamefont {Bruno}\ \bibnamefont
  {Bertini}}\ and\ \bibinfo {author} {\bibfnamefont {Lorenzo}\ \bibnamefont
  {Piroli}},\ }\bibfield  {title} {\enquote {\bibinfo {title} {Scrambling in
  random unitary circuits: {Exact} results},}\ }\href
  {https://link.aps.org/doi/10.1103/PhysRevB.102.064305} {\bibfield  {journal}
  {\bibinfo  {journal} {Phys. Rev. B}\ }\textbf {\bibinfo {volume} {102}},\
  \bibinfo {pages} {064305} (\bibinfo {year} {2020})}\BibitemShut {NoStop}%
\bibitem [{\citenamefont {Zhou}\ and\ \citenamefont
  {Harrow}(2022)}]{ZhouMaximal2022}%
  \BibitemOpen
  \bibfield  {author} {\bibinfo {author} {\bibfnamefont {Tianci}\ \bibnamefont
  {Zhou}}\ and\ \bibinfo {author} {\bibfnamefont {Aram~W.}\ \bibnamefont
  {Harrow}},\ }\bibfield  {title} {\enquote {\bibinfo {title} {Maximal
  entanglement velocity implies dual unitarity},}\ }\href
  {https://link.aps.org/doi/10.1103/PhysRevB.106.L201104} {\bibfield  {journal}
  {\bibinfo  {journal} {Phys. Rev. B}\ }\textbf {\bibinfo {volume} {106}},\
  \bibinfo {pages} {L201104} (\bibinfo {year} {2022})}\BibitemShut {NoStop}%
\bibitem [{\citenamefont {Bensa}\ and\ \citenamefont {{\v Z}nidari{\v
  c}}(2021)}]{bensa_fastest_2021}%
  \BibitemOpen
  \bibfield  {author} {\bibinfo {author} {\bibfnamefont {Ja{\v s}}\
  \bibnamefont {Bensa}}\ and\ \bibinfo {author} {\bibfnamefont {Marko}\
  \bibnamefont {{\v Z}nidari{\v c}}},\ }\bibfield  {title} {\enquote {\bibinfo
  {title} {Fastest {Local} {Entanglement} {Scrambler}, {Multistage}
  {Thermalization}, and a {Non}-{Hermitian} {Phantom}},}\ }\href
  {https://link.aps.org/doi/10.1103/PhysRevX.11.031019} {\bibfield  {journal}
  {\bibinfo  {journal} {Phys. Rev. X}\ }\textbf {\bibinfo {volume} {11}},\
  \bibinfo {pages} {031019} (\bibinfo {year} {2021})}\BibitemShut {NoStop}%
\bibitem [{\citenamefont {Alet}\ and\ \citenamefont
  {Laflorencie}(2018)}]{alet_many-body_2018}%
  \BibitemOpen
  \bibfield  {author} {\bibinfo {author} {\bibfnamefont {Fabien}\ \bibnamefont
  {Alet}}\ and\ \bibinfo {author} {\bibfnamefont {Nicolas}\ \bibnamefont
  {Laflorencie}},\ }\bibfield  {title} {\enquote {\bibinfo {title} {Many-body
  localization: {An} introduction and selected topics},}\ }\href
  {https://www.sciencedirect.com/science/article/pii/S163107051830032X}
  {\bibfield  {journal} {\bibinfo  {journal} {Comptes Rendus Physique}\
  }\textbf {\bibinfo {volume} {19}},\ \bibinfo {pages} {498--525} (\bibinfo
  {year} {2018})}\BibitemShut {NoStop}%
\bibitem [{\citenamefont {Foxen}\ \emph {et~al.}(2020)\citenamefont {Foxen},
  \citenamefont {Neill}, \citenamefont {Dunsworth}, \citenamefont {Roushan},
  \citenamefont {Chiaro}, \citenamefont {Megrant}, \citenamefont {Kelly},
  \citenamefont {Chen}, \citenamefont {Satzinger}, \citenamefont {Barends},
  \citenamefont {Arute}, \citenamefont {Arya}, \citenamefont {Babbush},
  \citenamefont {Bacon}, \citenamefont {Bardin}, \citenamefont {Boixo},
  \citenamefont {Buell}, \citenamefont {Burkett}, \citenamefont {Chen},
  \citenamefont {Collins}, \citenamefont {Farhi}, \citenamefont {Fowler},
  \citenamefont {Gidney}, \citenamefont {Giustina}, \citenamefont {Graff},
  \citenamefont {Harrigan}, \citenamefont {Huang}, \citenamefont {Isakov},
  \citenamefont {Jeffrey}, \citenamefont {Jiang}, \citenamefont {Kafri},
  \citenamefont {Kechedzhi}, \citenamefont {Klimov}, \citenamefont {Korotkov},
  \citenamefont {Kostritsa}, \citenamefont {Landhuis}, \citenamefont {Lucero},
  \citenamefont {McClean}, \citenamefont {McEwen}, \citenamefont {Mi},
  \citenamefont {Mohseni}, \citenamefont {Mutus}, \citenamefont {Naaman},
  \citenamefont {Neeley}, \citenamefont {Niu}, \citenamefont {Petukhov},
  \citenamefont {Quintana}, \citenamefont {Rubin}, \citenamefont {Sank},
  \citenamefont {Smelyanskiy}, \citenamefont {Vainsencher}, \citenamefont
  {White}, \citenamefont {Yao}, \citenamefont {Yeh}, \citenamefont {Zalcman},
  \citenamefont {Neven},\ and\ \citenamefont
  {Martinis}}]{Foxen2020Demonstrating}%
  \BibitemOpen
  \bibfield  {author} {\bibinfo {author} {\bibfnamefont {B.}~\bibnamefont
  {Foxen}}, \bibinfo {author} {\bibfnamefont {C.}~\bibnamefont {Neill}},
  \bibinfo {author} {\bibfnamefont {A.}~\bibnamefont {Dunsworth}}, \bibinfo
  {author} {\bibfnamefont {P.}~\bibnamefont {Roushan}}, \bibinfo {author}
  {\bibfnamefont {B.}~\bibnamefont {Chiaro}}, \bibinfo {author} {\bibfnamefont
  {A.}~\bibnamefont {Megrant}}, \bibinfo {author} {\bibfnamefont
  {J.}~\bibnamefont {Kelly}}, \bibinfo {author} {\bibfnamefont {Zijun}\
  \bibnamefont {Chen}}, \bibinfo {author} {\bibfnamefont {K.}~\bibnamefont
  {Satzinger}}, \bibinfo {author} {\bibfnamefont {R.}~\bibnamefont {Barends}},
  \bibinfo {author} {\bibfnamefont {F.}~\bibnamefont {Arute}}, \bibinfo
  {author} {\bibfnamefont {K.}~\bibnamefont {Arya}}, \bibinfo {author}
  {\bibfnamefont {R.}~\bibnamefont {Babbush}}, \bibinfo {author} {\bibfnamefont
  {D.}~\bibnamefont {Bacon}}, \bibinfo {author} {\bibfnamefont {J.~C.}\
  \bibnamefont {Bardin}}, \bibinfo {author} {\bibfnamefont {S.}~\bibnamefont
  {Boixo}}, \bibinfo {author} {\bibfnamefont {D.}~\bibnamefont {Buell}},
  \bibinfo {author} {\bibfnamefont {B.}~\bibnamefont {Burkett}}, \bibinfo
  {author} {\bibfnamefont {Yu}~\bibnamefont {Chen}}, \bibinfo {author}
  {\bibfnamefont {R.}~\bibnamefont {Collins}}, \bibinfo {author} {\bibfnamefont
  {E.}~\bibnamefont {Farhi}}, \bibinfo {author} {\bibfnamefont
  {A.}~\bibnamefont {Fowler}}, \bibinfo {author} {\bibfnamefont
  {C.}~\bibnamefont {Gidney}}, \bibinfo {author} {\bibfnamefont
  {M.}~\bibnamefont {Giustina}}, \bibinfo {author} {\bibfnamefont
  {R.}~\bibnamefont {Graff}}, \bibinfo {author} {\bibfnamefont
  {M.}~\bibnamefont {Harrigan}}, \bibinfo {author} {\bibfnamefont
  {T.}~\bibnamefont {Huang}}, \bibinfo {author} {\bibfnamefont {S.~V.}\
  \bibnamefont {Isakov}}, \bibinfo {author} {\bibfnamefont {E.}~\bibnamefont
  {Jeffrey}}, \bibinfo {author} {\bibfnamefont {Z.}~\bibnamefont {Jiang}},
  \bibinfo {author} {\bibfnamefont {D.}~\bibnamefont {Kafri}}, \bibinfo
  {author} {\bibfnamefont {K.}~\bibnamefont {Kechedzhi}}, \bibinfo {author}
  {\bibfnamefont {P.}~\bibnamefont {Klimov}}, \bibinfo {author} {\bibfnamefont
  {A.}~\bibnamefont {Korotkov}}, \bibinfo {author} {\bibfnamefont
  {F.}~\bibnamefont {Kostritsa}}, \bibinfo {author} {\bibfnamefont
  {D.}~\bibnamefont {Landhuis}}, \bibinfo {author} {\bibfnamefont
  {E.}~\bibnamefont {Lucero}}, \bibinfo {author} {\bibfnamefont
  {J.}~\bibnamefont {McClean}}, \bibinfo {author} {\bibfnamefont
  {M.}~\bibnamefont {McEwen}}, \bibinfo {author} {\bibfnamefont
  {X.}~\bibnamefont {Mi}}, \bibinfo {author} {\bibfnamefont {M.}~\bibnamefont
  {Mohseni}}, \bibinfo {author} {\bibfnamefont {J.~Y.}\ \bibnamefont {Mutus}},
  \bibinfo {author} {\bibfnamefont {O.}~\bibnamefont {Naaman}}, \bibinfo
  {author} {\bibfnamefont {M.}~\bibnamefont {Neeley}}, \bibinfo {author}
  {\bibfnamefont {M.}~\bibnamefont {Niu}}, \bibinfo {author} {\bibfnamefont
  {A.}~\bibnamefont {Petukhov}}, \bibinfo {author} {\bibfnamefont
  {C.}~\bibnamefont {Quintana}}, \bibinfo {author} {\bibfnamefont
  {N.}~\bibnamefont {Rubin}}, \bibinfo {author} {\bibfnamefont
  {D.}~\bibnamefont {Sank}}, \bibinfo {author} {\bibfnamefont {V.}~\bibnamefont
  {Smelyanskiy}}, \bibinfo {author} {\bibfnamefont {A.}~\bibnamefont
  {Vainsencher}}, \bibinfo {author} {\bibfnamefont {T.~C.}\ \bibnamefont
  {White}}, \bibinfo {author} {\bibfnamefont {Z.}~\bibnamefont {Yao}}, \bibinfo
  {author} {\bibfnamefont {P.}~\bibnamefont {Yeh}}, \bibinfo {author}
  {\bibfnamefont {A.}~\bibnamefont {Zalcman}}, \bibinfo {author} {\bibfnamefont
  {H.}~\bibnamefont {Neven}}, \ and\ \bibinfo {author} {\bibfnamefont {J.~M.}\
  \bibnamefont {Martinis}} (\bibinfo {collaboration} {Google AI Quantum}),\
  }\bibfield  {title} {\enquote {\bibinfo {title} {Demonstrating a continuous
  set of two-qubit gates for near-term quantum algorithms},}\ }\href
  {https://link.aps.org/doi/10.1103/PhysRevLett.125.120504} {\bibfield
  {journal} {\bibinfo  {journal} {Phys. Rev. Lett.}\ }\textbf {\bibinfo
  {volume} {125}},\ \bibinfo {pages} {120504} (\bibinfo {year}
  {2020})}\BibitemShut {NoStop}%
\bibitem [{\citenamefont {Garratt}\ and\ \citenamefont
  {Chalker}(2021)}]{garratt_many-body_2021}%
  \BibitemOpen
  \bibfield  {author} {\bibinfo {author} {\bibfnamefont {S.~J.}\ \bibnamefont
  {Garratt}}\ and\ \bibinfo {author} {\bibfnamefont {J.~T.}\ \bibnamefont
  {Chalker}},\ }\bibfield  {title} {\enquote {\bibinfo {title} {Many-body
  delocalization as symmetry breaking},}\ }\href
  {https://link.aps.org/doi/10.1103/PhysRevLett.127.026802} {\bibfield
  {journal} {\bibinfo  {journal} {Phys. Rev. Lett.}\ }\textbf {\bibinfo
  {volume} {127}},\ \bibinfo {pages} {026802} (\bibinfo {year}
  {2021})}\BibitemShut {NoStop}%
\bibitem [{\citenamefont {Bertini}\ \emph
  {et~al.}(2019{\natexlab{c}})\citenamefont {Bertini}, \citenamefont {Kos},\
  and\ \citenamefont {Prosen}}]{Bertini2019Entanglement}%
  \BibitemOpen
  \bibfield  {author} {\bibinfo {author} {\bibfnamefont {Bruno}\ \bibnamefont
  {Bertini}}, \bibinfo {author} {\bibfnamefont {Pavel}\ \bibnamefont {Kos}}, \
  and\ \bibinfo {author} {\bibfnamefont {Tomaz}\ \bibnamefont {Prosen}},\
  }\bibfield  {title} {\enquote {\bibinfo {title} {Entanglement spreading in a
  minimal model of maximal many-body quantum chaos},}\ }\href
  {https://link.aps.org/doi/10.1103/PhysRevX.9.021033} {\bibfield  {journal}
  {\bibinfo  {journal} {Phys. Rev. X}\ }\textbf {\bibinfo {volume} {9}},\
  \bibinfo {pages} {021033} (\bibinfo {year} {2019}{\natexlab{c}})}\BibitemShut
  {NoStop}%
\bibitem [{\citenamefont {Akila}\ \emph {et~al.}(2016)\citenamefont {Akila},
  \citenamefont {Waltner}, \citenamefont {Gutkin},\ and\ \citenamefont
  {Guhr}}]{Akila_2016}%
  \BibitemOpen
  \bibfield  {author} {\bibinfo {author} {\bibfnamefont {M}~\bibnamefont
  {Akila}}, \bibinfo {author} {\bibfnamefont {D}~\bibnamefont {Waltner}},
  \bibinfo {author} {\bibfnamefont {B}~\bibnamefont {Gutkin}}, \ and\ \bibinfo
  {author} {\bibfnamefont {T}~\bibnamefont {Guhr}},\ }\bibfield  {title}
  {\enquote {\bibinfo {title} {Particle-time duality in the kicked ising spin
  chain},}\ }\href {https://dx.doi.org/10.1088/1751-8113/49/37/375101}
  {\bibfield  {journal} {\bibinfo  {journal} {Journal of Physics A:
  Mathematical and Theoretical}\ }\textbf {\bibinfo {volume} {49}},\ \bibinfo
  {pages} {375101} (\bibinfo {year} {2016})}\BibitemShut {NoStop}%
\bibitem [{\citenamefont {Rampp}\ \emph {et~al.}(2023)\citenamefont {Rampp},
  \citenamefont {Moessner},\ and\ \citenamefont {Claeys}}]{rampp_dual_2023}%
  \BibitemOpen
  \bibfield  {author} {\bibinfo {author} {\bibfnamefont {Michael~A.}\
  \bibnamefont {Rampp}}, \bibinfo {author} {\bibfnamefont {Roderich}\
  \bibnamefont {Moessner}}, \ and\ \bibinfo {author} {\bibfnamefont
  {Pieter~W.}\ \bibnamefont {Claeys}},\ }\bibfield  {title} {\enquote {\bibinfo
  {title} {From {Dual} {Unitarity} to {Generic} {Quantum} {Operator}
  {Spreading}},}\ }\href
  {https://link.aps.org/doi/10.1103/PhysRevLett.130.130402} {\bibfield
  {journal} {\bibinfo  {journal} {Phys. Rev. Lett.}\ }\textbf {\bibinfo
  {volume} {130}},\ \bibinfo {pages} {130402} (\bibinfo {year}
  {2023})}\BibitemShut {NoStop}%
\bibitem [{\citenamefont {Pal}\ and\ \citenamefont
  {Huse}(2010)}]{pal_many-body_2010}%
  \BibitemOpen
  \bibfield  {author} {\bibinfo {author} {\bibfnamefont {Arijeet}\ \bibnamefont
  {Pal}}\ and\ \bibinfo {author} {\bibfnamefont {David~A.}\ \bibnamefont
  {Huse}},\ }\bibfield  {title} {\enquote {\bibinfo {title} {Many-body
  localization phase transition},}\ }\href
  {https://link.aps.org/doi/10.1103/PhysRevB.82.174411} {\bibfield  {journal}
  {\bibinfo  {journal} {Phys. Rev. B}\ }\textbf {\bibinfo {volume} {82}},\
  \bibinfo {pages} {174411} (\bibinfo {year} {2010})}\BibitemShut {NoStop}%
\bibitem [{ata()}]{atas_distribution_2013}%
  \BibitemOpen
  \bibfield  {title} {\enquote {\bibinfo {title} {Distribution of the {Ratio}
  of {Consecutive} {Level} {Spacings} in {Random} {Matrix} {Ensembles}},}\
  }\href {https://link.aps.org/doi/10.1103/PhysRevLett.110.084101} {\ \textbf
  {\bibinfo {volume} {110}},\ \bibinfo {pages} {084101}}\BibitemShut {NoStop}%
\bibitem [{\citenamefont {Page}(1993)}]{page_average_1993}%
  \BibitemOpen
  \bibfield  {author} {\bibinfo {author} {\bibfnamefont {Don~N.}\ \bibnamefont
  {Page}},\ }\bibfield  {title} {\enquote {\bibinfo {title} {Average entropy of
  a subsystem},}\ }\href {https://link.aps.org/doi/10.1103/PhysRevLett.71.1291}
  {\bibfield  {journal} {\bibinfo  {journal} {Phys. Rev. Lett.}\ }\textbf
  {\bibinfo {volume} {71}},\ \bibinfo {pages} {1291--1294} (\bibinfo {year}
  {1993})}\BibitemShut {NoStop}%
\bibitem [{\citenamefont {Khemani}\ \emph {et~al.}(2017)\citenamefont
  {Khemani}, \citenamefont {Lim}, \citenamefont {Sheng},\ and\ \citenamefont
  {Huse}}]{khemani_critical_2017}%
  \BibitemOpen
  \bibfield  {author} {\bibinfo {author} {\bibfnamefont {Vedika}\ \bibnamefont
  {Khemani}}, \bibinfo {author} {\bibfnamefont {S.~P.}\ \bibnamefont {Lim}},
  \bibinfo {author} {\bibfnamefont {D.~N.}\ \bibnamefont {Sheng}}, \ and\
  \bibinfo {author} {\bibfnamefont {David~A.}\ \bibnamefont {Huse}},\
  }\bibfield  {title} {\enquote {\bibinfo {title} {Critical properties of the
  many-body localization transition},}\ }\href
  {https://link.aps.org/doi/10.1103/PhysRevX.7.021013} {\bibfield  {journal}
  {\bibinfo  {journal} {Phys. Rev. X}\ }\textbf {\bibinfo {volume} {7}},\
  \bibinfo {pages} {021013} (\bibinfo {year} {2017})}\BibitemShut {NoStop}%
\bibitem [{yu_()}]{yu_bimodal_2016}%
  \BibitemOpen
  \bibfield  {title} {\enquote {\bibinfo {title} {Bimodal entanglement entropy
  distribution in the many-body localization transition},}\ }\href
  {https://link.aps.org/doi/10.1103/PhysRevB.94.184202} {\bibfield  {journal}
  {\bibinfo  {journal} {Phys. Rev. B}\ }\textbf {\bibinfo {volume} {94}},\
  \bibinfo {pages} {184202}}\BibitemShut {NoStop}%
\bibitem [{\citenamefont {Luitz}(2021)}]{luitz_polynomial_2021}%
  \BibitemOpen
  \bibfield  {author} {\bibinfo {author} {\bibfnamefont {David~J.}\
  \bibnamefont {Luitz}},\ }\bibfield  {title} {\enquote {\bibinfo {title}
  {Polynomial filter diagonalization of large {Floquet} unitary operators},}\
  }\href {https://scipost.org/10.21468/SciPostPhys.11.2.021} {\bibfield
  {journal} {\bibinfo  {journal} {SciPost Physics}\ }\textbf {\bibinfo {volume}
  {11}},\ \bibinfo {pages} {021} (\bibinfo {year} {2021})}\BibitemShut
  {NoStop}%
\bibitem [{\citenamefont {Long}\ \emph {et~al.}(2023)\citenamefont {Long},
  \citenamefont {Crowley}, \citenamefont {Khemani},\ and\ \citenamefont
  {Chandran}}]{long_phenomenology_2022}%
  \BibitemOpen
  \bibfield  {author} {\bibinfo {author} {\bibfnamefont {David~M.}\
  \bibnamefont {Long}}, \bibinfo {author} {\bibfnamefont {Philip J.~D.}\
  \bibnamefont {Crowley}}, \bibinfo {author} {\bibfnamefont {Vedika}\
  \bibnamefont {Khemani}}, \ and\ \bibinfo {author} {\bibfnamefont {Anushya}\
  \bibnamefont {Chandran}},\ }\bibfield  {title} {\enquote {\bibinfo {title}
  {Phenomenology of the prethermal many-body localized regime},}\ }\href
  {https://link.aps.org/doi/10.1103/PhysRevLett.131.106301} {\bibfield
  {journal} {\bibinfo  {journal} {Phys. Rev. Lett.}\ }\textbf {\bibinfo
  {volume} {131}},\ \bibinfo {pages} {106301} (\bibinfo {year}
  {2023})}\BibitemShut {NoStop}%
\bibitem [{\citenamefont {Luitz}\ \emph {et~al.}(2016)\citenamefont {Luitz},
  \citenamefont {Laflorencie},\ and\ \citenamefont
  {Alet}}]{luitz_extended_2016}%
  \BibitemOpen
  \bibfield  {author} {\bibinfo {author} {\bibfnamefont {David~J.}\
  \bibnamefont {Luitz}}, \bibinfo {author} {\bibfnamefont {Nicolas}\
  \bibnamefont {Laflorencie}}, \ and\ \bibinfo {author} {\bibfnamefont
  {Fabien}\ \bibnamefont {Alet}},\ }\bibfield  {title} {\enquote {\bibinfo
  {title} {Extended slow dynamical regime close to the many-body localization
  transition},}\ }\href {https://link.aps.org/doi/10.1103/PhysRevB.93.060201}
  {\bibfield  {journal} {\bibinfo  {journal} {Phys. Rev. B}\ }\textbf {\bibinfo
  {volume} {93}},\ \bibinfo {pages} {060201(R)} (\bibinfo {year}
  {2016})}\BibitemShut {NoStop}%
\bibitem [{\citenamefont {Lezama}\ \emph {et~al.}(2019)\citenamefont {Lezama},
  \citenamefont {Bera},\ and\ \citenamefont
  {Bardarson}}]{lezama_apparent_2019}%
  \BibitemOpen
  \bibfield  {author} {\bibinfo {author} {\bibfnamefont {Talía L.~M.}\
  \bibnamefont {Lezama}}, \bibinfo {author} {\bibfnamefont {Soumya}\
  \bibnamefont {Bera}}, \ and\ \bibinfo {author} {\bibfnamefont {Jens~H.}\
  \bibnamefont {Bardarson}},\ }\bibfield  {title} {\enquote {\bibinfo {title}
  {Apparent slow dynamics in the ergodic phase of a driven many-body localized
  system without extensive conserved quantities},}\ }\href
  {https://link.aps.org/doi/10.1103/PhysRevB.99.161106} {\bibfield  {journal}
  {\bibinfo  {journal} {Phys. Rev. B}\ }\textbf {\bibinfo {volume} {99}},\
  \bibinfo {pages} {161106(R)} (\bibinfo {year} {2019})}\BibitemShut {NoStop}%
\bibitem [{\citenamefont {Lezama}\ \emph {et~al.}(2021)\citenamefont {Lezama},
  \citenamefont {Torres-Herrera}, \citenamefont {Perez-Bernal}, \citenamefont
  {BarLev},\ and\ \citenamefont {Santos}}]{lezama_equilibration_2021}%
  \BibitemOpen
  \bibfield  {author} {\bibinfo {author} {\bibfnamefont {Talía L.~M.}\
  \bibnamefont {Lezama}}, \bibinfo {author} {\bibfnamefont {E.~J.}\
  \bibnamefont {Torres-Herrera}}, \bibinfo {author} {\bibfnamefont
  {F.}~\bibnamefont {Perez-Bernal}}, \bibinfo {author} {\bibfnamefont
  {Yevgeny}\ \bibnamefont {BarLev}}, \ and\ \bibinfo {author} {\bibfnamefont
  {Lea~F.}\ \bibnamefont {Santos}},\ }\bibfield  {title} {\enquote {\bibinfo
  {title} {Equilibration time in many-body quantum systems},}\ }\href
  {https://link.aps.org/doi/10.1103/PhysRevB.104.085117} {\bibfield  {journal}
  {\bibinfo  {journal} {Phys. Rev. B}\ }\textbf {\bibinfo {volume} {104}},\
  \bibinfo {pages} {085117} (\bibinfo {year} {2021})}\BibitemShut {NoStop}%
\bibitem [{\citenamefont {Kjall}\ \emph {et~al.}(2014)\citenamefont {Kjall},
  \citenamefont {Bardarson},\ and\ \citenamefont
  {Pollmann}}]{kjall_many-body_2014}%
  \BibitemOpen
  \bibfield  {author} {\bibinfo {author} {\bibfnamefont {J.~A.}\ \bibnamefont
  {Kjall}}, \bibinfo {author} {\bibfnamefont {J.~H.}\ \bibnamefont
  {Bardarson}}, \ and\ \bibinfo {author} {\bibfnamefont {Frank}\ \bibnamefont
  {Pollmann}},\ }\bibfield  {title} {\enquote {\bibinfo {title} {Many-{Body}
  {Localization} in a {Disordered} {Quantum} {Ising} {Chain}},}\ }\href
  {https://link.aps.org/doi/10.1103/PhysRevLett.113.107204} {\bibfield
  {journal} {\bibinfo  {journal} {Phys. Rev. Lett.}\ }\textbf {\bibinfo
  {volume} {113}},\ \bibinfo {pages} {107204} (\bibinfo {year}
  {2014})}\BibitemShut {NoStop}%
\bibitem [{\citenamefont {Ponte}\ \emph {et~al.}(2015)\citenamefont {Ponte},
  \citenamefont {Papi\ifmmode~\acute{c}\else \'{c}\fi{}}, \citenamefont
  {Huveneers},\ and\ \citenamefont {Abanin}}]{Ponte2015Many-body}%
  \BibitemOpen
  \bibfield  {author} {\bibinfo {author} {\bibfnamefont {Pedro}\ \bibnamefont
  {Ponte}}, \bibinfo {author} {\bibfnamefont {Z.}~\bibnamefont
  {Papi\ifmmode~\acute{c}\else \'{c}\fi{}}}, \bibinfo {author} {\bibfnamefont
  {Fran\ifmmode \mbox{\c{c}}\else~\c{c}\fi{}ois}\ \bibnamefont {Huveneers}}, \
  and\ \bibinfo {author} {\bibfnamefont {Dmitry~A.}\ \bibnamefont {Abanin}},\
  }\bibfield  {title} {\enquote {\bibinfo {title} {Many-body localization in
  periodically driven systems},}\ }\href
  {https://link.aps.org/doi/10.1103/PhysRevLett.114.140401} {\bibfield
  {journal} {\bibinfo  {journal} {Phys. Rev. Lett.}\ }\textbf {\bibinfo
  {volume} {114}},\ \bibinfo {pages} {140401} (\bibinfo {year}
  {2015})}\BibitemShut {NoStop}%
\bibitem [{\citenamefont {Kj\"all}\ \emph {et~al.}(2014)\citenamefont
  {Kj\"all}, \citenamefont {Bardarson},\ and\ \citenamefont
  {Pollmann}}]{Kjaell2014Many-body}%
  \BibitemOpen
  \bibfield  {author} {\bibinfo {author} {\bibfnamefont {Jonas~A.}\
  \bibnamefont {Kj\"all}}, \bibinfo {author} {\bibfnamefont {Jens~H.}\
  \bibnamefont {Bardarson}}, \ and\ \bibinfo {author} {\bibfnamefont {Frank}\
  \bibnamefont {Pollmann}},\ }\bibfield  {title} {\enquote {\bibinfo {title}
  {Many-body localization in a disordered quantum ising chain},}\ }\href
  {https://link.aps.org/doi/10.1103/PhysRevLett.113.107204} {\bibfield
  {journal} {\bibinfo  {journal} {Phys. Rev. Lett.}\ }\textbf {\bibinfo
  {volume} {113}},\ \bibinfo {pages} {107204} (\bibinfo {year}
  {2014})}\BibitemShut {NoStop}%
\bibitem [{\citenamefont {Developers}(2023)}]{cirq_developers_2023_8161252}%
  \BibitemOpen
  \bibfield  {author} {\bibinfo {author} {\bibfnamefont {Cirq}\ \bibnamefont
  {Developers}},\ }\href {\doibase 10.5281/zenodo.8161252} {\enquote {\bibinfo
  {title} {Cirq},}\ } (\bibinfo {year} {2023}),\ \bibinfo {note} {{See full
  list of authors on Github: https://github
  .com/quantumlib/Cirq/graphs/contributors}}\BibitemShut {NoStop}%
\bibitem [{\citenamefont {Krajewski}\ \emph {et~al.}(2023)\citenamefont
  {Krajewski}, \citenamefont {Vidmar}, \citenamefont {Bonca},\ and\
  \citenamefont {Mierzejewski}}]{krajewski_strongly_2023}%
  \BibitemOpen
  \bibfield  {author} {\bibinfo {author} {\bibfnamefont {B.}~\bibnamefont
  {Krajewski}}, \bibinfo {author} {\bibfnamefont {L.}~\bibnamefont {Vidmar}},
  \bibinfo {author} {\bibfnamefont {J.}~\bibnamefont {Bonca}}, \ and\ \bibinfo
  {author} {\bibfnamefont {M.}~\bibnamefont {Mierzejewski}},\ }\bibfield
  {title} {\enquote {\bibinfo {title} {Strongly disordered {Anderson} insulator
  chains with generic two-body interaction},}\ }\href {\doibase
  10.1103/PhysRevB.108.064203} {\bibfield  {journal} {\bibinfo  {journal}
  {Phys. Rev. B}\ }\textbf {\bibinfo {volume} {108}},\ \bibinfo {pages}
  {064203} (\bibinfo {year} {2023})}\BibitemShut {NoStop}%
\end{thebibliography}%

\newpage
\clearpage

\appendix

\section{Gate operator entanglement, SWAP and CNOT lines}\label{sec:opee_swap_cnot}

\begin{figure}[h]
    \centering
    \includegraphics{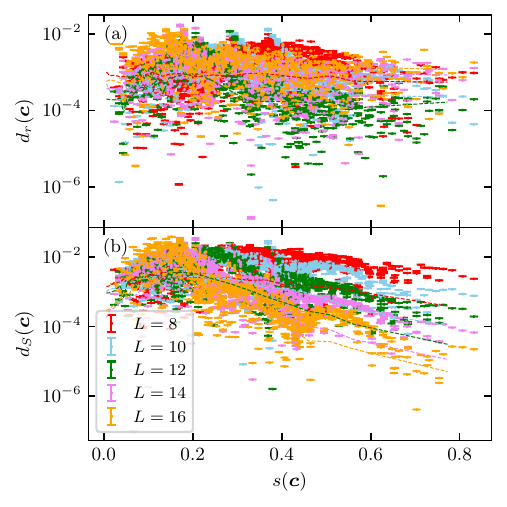}
    \caption{Distance of (a) the gap ratio $r(\vec{c},L)$ and (b) entanglement entropy $S(\vec{c},L)$ to the reference values $r(\vec{c}_S,L)$ and $S(\vec{c}_S,L)$ on the SWAP line. Dashed lines are error bars of $r(\vec{c}_S,L)$ and $S(\vec{c}_S,L)$. The SWAP reference points are computed by using spline interpolation of the curves shown in Fig.~\ref{fig:gapratio_and_EE}. The points are the same data as shown in Fig.~\ref{fig:gap_ratio_EE_all}.}
    \label{fig:gap_ratio_EE_diff}
\end{figure}

Throughout this work, we assume that the results for the SWAP, iSWAP and CNOT line can be extrapolated to any other set of $\vec{c}=(c_1,c_2,c_3)$ via the equivalence through the gate operator entanglement $s(\vec{c})$. Inspired by Ref.~\cite{rampp_dual_2023}, where the authors found that $s(\vec{c})$ plays a leading role in generic thermalization, we aim to apply these ideas to the weakly interacting regime. In this section, we provide quantitative tests for this assumption by computing the gap ratio and eigenstate operator entanglement at any point $(c_1,c_2,c_3)$, namely $r(\vec{c},L)$ and $S(\vec{c},L)$, and the difference
\begin{align}
d_r(\vec{c})&=|r(\vec{c},L)-r(\vec{c}_S,L)|\\
d_S(\vec{c})&=|S(\vec{c},L)-S(\vec{c}_S,L)|/S_{Page},
\end{align}
where $r(\vec{c}_S,L)$ and $S(\vec{c}_S,L)$ are the gap ratio and eigenstate entanglement entropy for the reference choice ${\vec{c}_S=(c_x,c_x,c_x)}$ on the SWAP line and the same gate operator entanglement entropy $s(\vec{c})=s(\vec{c}_S)$. Our results are shown in Fig.~\ref{fig:gap_ratio_EE_diff}.
As a reference, we compare these results with the effects of the statistical error of $r(\vec{c}_S,L)$ and $S(\vec{c}_S,L)$ due to the disorder average, indicated by dashed lines. When a point $d_{r}(\vec{c})$ or $d_S(\vec{c})$ is below the corresponding error bar line, then the difference of the results for the parameter $\vec{c}$ and the reference point $\vec{c}_S$ are the same within statistical errors. Visually, there is a large fraction of $d_{r}(\vec{c})$ or $d_S(\vec{c})$ above the error bar line for all system sizes. 
However, the difference in absolute value is small enough, such that extrapolating the results from the SWAP, iSWAP and CNOT line to other values of $\vec{c}$ is a fair assumption for the current setup and system sizes.  
\section{Comparison with scaling of integrability breaking perturbations}\label{app:integrabbreaking}
\begin{figure}
    \centering
    \includegraphics{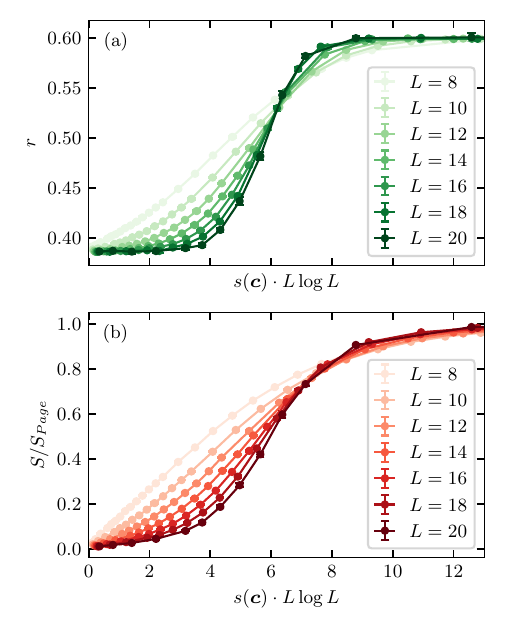}
    \caption{(a) Gap ratio and (b) entanglement entropy along the SWAP line for system size $L=8-20$. The two-qubit gate operator entanglement $s(\vec{c})$ is re-scaled by $L\log L$. The data set is the same as shown in Fig.~\ref{fig:gapratio_and_EE}. These visible crossings suggest a drifting critical value $s(\vec{c})_{crit} \propto 1/L$, compatible with an integrability-breaking phenomenon~(see App.~\ref{app:integrabbreaking)}).}
    \label{fig:gap_ratio_rescaled}
\end{figure}
It has been recently shown that non-interacting spin systems with small perturbative interaction undergo a Fock space type delocalization ``transition" that marks the onset of quantum chaos \cite{bulchandani_onset_2022}. Taking $\epsilon$ as the integrability breaking parameter, the value $\epsilon_c$ denotes the onset of chaos scaling as $\epsilon_c \sim (L\log L)^{-1}$ with increasing system size. In Fig.~\ref{fig:gap_ratio_rescaled} we test such scaling for $s(\vec{c})$ for both gap ratio and eigenstate entanglement entropy. Although the available system sizes do not allow us to discern the $\log L$ component, the linear scaling is clearly visible. From this perspective, the  MBL-thermalization crossover appears to have a similar scaling as an integrability-breaking phenomenon for finite system sizes.

\end{document}